\documentclass[aps,prc,preprint,showpacs]{revtex4}
\usepackage{amsfonts}
\usepackage{amssymb}
\usepackage{graphics}
\usepackage{graphicx}
\usepackage{color}
\usepackage{slashed}
\usepackage{enumerate}
\begin{document}

\title{Asymmetry of the neutrino mean free path in hot neutron matter under strong magnetic fields}

\author{Julio Torres Pati\~no$^{1}$, Eduardo Bauer$^{1,2}$ and Isaac Vida\~na$^3$}
\affiliation{$^1$ IFLP, CCT-La Plata, CONICET. Argentina}
\affiliation{$^2$Facultad de Ciencias Astron\'omicas y Geof\'{\i}sicas, Universidad Nacional de La Plata, Argentina}
\affiliation{$^3$Istituto  Nazionale  di  Fisica  Nucleare, Dipartimento  di  Fisica,  Universit$\grave{a}$  di  Catania, Via  Santa  Sofia  64,  I-95123,  Catania,  Italy}


\begin{abstract}
The neutrino mean free path in neutron matter under a strong magnetic field is evaluated for the inelastic scattering reaction and studied as a function of the neutron matter density in the range $0.05 \leq \rho \leq 0.4$ fm$^{-3}$ for several temperatures up to 30 MeV and magnetic field strengths B=0 G, $10^{18}$ G and $2.5\times 10^{18}$ G.
Polarized neutron matter is described within the non--relativistic Brueckner--Hartree--Fock (BHF) approach using the Argonne V18 nucleon-nucleon potential supplemented with the Urbana IX three-nucleon force. Explicit expressions of the cross section per unit volume for the scattering of a neutrino with a spin up or spin down neutron are derived from the Fermi Golden rule. Our results show that the mean free path depends strongly on the angle of the incoming neutrino, leading to an asymmetry in this quantity. This asymmetry depends on the magnetic field intensity and on the density, but it is rather independent of the temperature. For a density of $0.16$ fm$^{-3}$ at a temperature T$=30$MeV, the asymmetry in the mean free path is found to be of $\sim 15\%$ for B=$10^{18}$G and $\sim 38\%$ for B=$2.5 \times 10^{18}$G.
\end{abstract}


\maketitle

\section{Introduction}
\label{int}

Neutrinos play a crucial role in the physics of supernova explosions \cite{bethe,janka,burrows}, during the early evolution of compact stellar remnants \cite{burrowsb, jankab}, in neutron star cooling \cite{Sh96, Ya04}, and in neutron star mergers \cite{perego1,perego2,perego3}. A large number of neutrinos are produced by electron capture processes during the gravitational collapse of the core of a massive star. Most of the initial gravitational binding energy is stored and released by the neutrinos. In the early stages following the formation of a neutron star the neutrino mean free path $\lambda$ decreases and, above a critical value of the density, becomes smaller than the stellar radius. Under these conditions neutrinos are {\it trapped} in the star. Neutrino trapping has a strong influence on the overall {\it stiffness} of the equation of state (EoS) of dense matter \cite{bombaci1,bombaci2}, being the physical conditions of hot and lepton-rich neutron stars substantially different from those of the cold and deleptonized ones. The cooling of a newly born hot neutron star is driven first by the neutrino emission from the interior. There are several neutrino emission processes that contribute to the cooling of neutron stars. These include among others, the direct and modified URCA processes, bremsstrahlung or Cooper pair formation, which operates only when the temperature of the star drops below the critical temperature for neutron superfluidity or proton superconductivity. Neutrino cross sections and emissivity are fundamental inputs for supernova simulations and cooling calculations. These quantities can be substantially affected by the presence of strong magnetic fields in the neutron stars. In the case of the so-called magnetars, the magnetic field intensity can reach values up to  $10^{14}-10^{15}$ G at the star surface and it can grow by several orders of magnitude in its dense interior \cite{Du92}.
The emission of neutrinos, for instance, is expected to be asymmetric ({\it i.e.,} to depend on the direction of the neutrino) under the presence of a strong magnetic field.

The asymmetrical emission of neutrinos has been suggested as a possible mechanism to explain the so-called ``pulsar kick problem": the observation that pulsars do not move with the velocity of its progenitor star, but rather with a substantially greater speed. Although an asymmetry as small as $\sim 1\%$ would be enough to explain the
pulsar movement, this mechanism has been questioned as the (unique) source for the ``pulsar kick"
(see for instance~\cite{Sa08}). Other possible explanatory mechanisms include: an asymmetry in the gravitational collapse of the progenitor, acceleration due to the pulsar electromagnetic radiation or the evolution of binary system which may produce rapidly moving pulsars. The asymmetrical emission of neutrinos can have different origins. Neutrino
oscillation can be altered by the magnetic field, resulting in an anisotropy in the momentum of
the outgoing neutrinos \cite{Ku96}. Parity violating can also induce an asymmetry on the neutrino emission when
multiple--scattering of neutrinos in slightly polarized neutrons is taking into account \cite{Ho98,Ka06}.
Here we are particularly interested in this last mechanism, which on practice results from the addition of a modified differential cross section plus the cumulative effect of multiple--scattering. In this case, two ingredients are important: the differential cross section and the neutrino mean free path. Note that in the absence of a magnetic field
the non--relativistic elastic differential cross section of neutrinos with
neutron matter can be written as,
\begin{equation}
\frac{d \sigma}{d \Omega} = \frac{G^{2}_{F} E^{2}_{\nu}}{4 \pi^{2}} (C^{2}_{A} (3 - \cos \theta)+
C^{2}_{V} (1 + \cos \theta)) \, ,
\label{eq:diffxs}
\end{equation}
where $G_{F}$ is the Fermi coupling constant and $\theta$ is the scattering angle.
Even though the differential cross section is not uniform, in the absence of a preference spatial axis, the average emission of
neutrinos from the whole neutron star would be isotropic. However, the presence of an uniform magnetic
field modifies this expression and produces an asymmetry in the neutrino emission.

The second ingredient, the neutrino mean free path in dense matter (defined as the inverse of the total neutrino cross section per unit volume) has been studied in the absence of a magnetic field by many authors using various approximation schemes and various models of the trapping environment  (see {\it e.g.}
Refs\ \cite{tubbs,sawyer,iwamoto,backman,haensel87,Re99,Ho91,Re97,Re98,ReddyT,Bu98,Na99,Sh03,Ma03} and references therein). The behavior of neutrinos in dense matter under the presence of strong magnetic fields has been also considered in the literature \cite{Be96,Ku96,Ho98,Bai99,Ar99,Ch02,An03,Ho06,Sa08,Pe09a,Pe09b}. However, the asymmetry on the neutrino emission, due to the breaking of the isotropy by the field, has not been discussed much.

The scope of the present work is to analyze the effect of a strong magnetic field on the mean free path of neutrinos  in hot neutron matter focussing, in particular, on the asymmetry on the neutrino emission induced  by the presence of the field.
In neutron matter the two dominant mechanisms contributing to the neutrino mean free path are the {\it scattering} of the neutrino with a neutron and the {\it absorption} of the neutrino by the neutron producing a proton and an electron in the final state.  In this work, however, we will restrict ourselves to the first one of these mechanisms. The interested reader is referred {\it e.g.,} to~\cite{Re99} for a complete description of all possible reactions involving neutrinos.

In particular, we derive explicit expressions of the neutrino cross section per unit volume for the scattering of a neutrino with a spin up or spin down neutron. The description of  polarized neutron matter is made within the non--relativistic Brueckner--Hartree--Fock (BHF) approach using the Argonne V18 \cite{argonne} nucleon-nucleon potential supplemented with the Urbana IX \cite{urbana} three-nucleon force.

The paper is organized as follows. In Section~\ref{ncs}, we discuss the inelastic scattering neutrino cross section with polarized neutrons. Starting from the Fermi Golden rule, we develop expressions for the total neutrino cross section, taking the non--relativistic limit to be consistent with our EoS--model. In Section~\ref{Results}, we discuss some results, where we start with the properties of polarized neutron matter, we also discuss some general properties of the neutrino mean free path and then we show the asymmetry in this quantity. Finally, in Section~\ref{Summary} a summary, the main conclusions and future perspectives are given.

%
\newpage
\section{The Neutrino Cross Section}
\label{ncs}

\begin{figure}[t]
\begin{center}
    \includegraphics[width = 0.45 \textwidth]{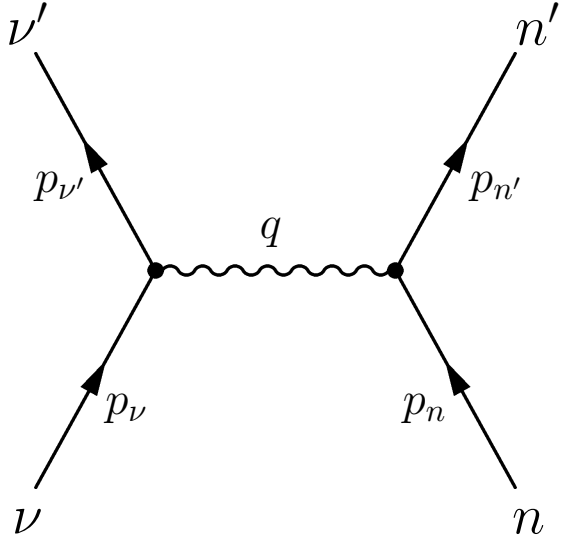}
\vskip 2mm
\caption{The lowest order Feynman diagram for the scattering reaction $\nu + n \to \nu' + n'$. The quantities $p_i$ and $q$ denote, respectively, the four--momentum of the involved particles and the
corresponding four--momentum transfer by the interaction.}
\label{figme1}
\end{center}
\end{figure}

In this section we derive the expression for the neutrino total cross section per unit volume  in hot neutron matter under the presence of a strong constant magnetic field. As it has already been said in the introduction, in this work we restrict ourselves to the neutrino scattering process,
\begin{equation}
\nu + n \to \nu' + n'  \ ,
\end{equation}
denoting $\nu$ and $n$ ($\nu'$ and $n'$) the incoming (outgoing) neutrino and neutron, respectively.
We note here that in this work neutrinos are considered massless. Fig.\ \ref{figme1} shows the lowest order Feynman diagram contributing to this reaction. Using the Fermi Golden Rule (see {\it e.g.} \cite{Joach}), we can write down the contribution of this reaction to the total cross section per unit volume simply as:
\begin{eqnarray}
\frac{\sigma(p_{\nu})}{V} & = & \int\frac{d {\vec p_{\nu'}}}{(2 \pi)^{3}} \int\frac{d {\vec p_{n}}}{(2 \pi)^{3}}
\int\frac{d {\vec p_{n'}}}{(2 \pi)^{3}} (2 \pi)^{4} \delta^{(4)}(p_{\nu}+p_{n}-p_{\nu'}-p_{n'}) \nonumber \\
&& \times f_{n}({\vec p_{n}}, T) (1-f_{n'}({\vec p_{n'}}, T)) \frac{| {\cal M}_{\nu' n',\nu n} |^{2}}
{2^{4} E_{\nu} E_{\nu'} E_{n} E_{n'}} \ ,
\label{cross1}
\end{eqnarray}
where $p_{i}=(E_{i},{\vec p}_{i})$ is the four-momentum of particle $i$, ${\cal M}_{\nu' n',\nu n}$ is the so-called M{\o}ller invariant transition matrix, which we discuss below, and $f_{i}({\vec p}_{i}, T)$ is the particle distribution function, which in thermal equilibrium is given by the Fermi--Dirac one,
\begin{equation}
f_{i}({\vec p}_{i}, T)=\frac{1}{1+\exp[(E_{i}({\vec p}_{i}, T)-\mu_i(T))/T]} \ ,
\label{fdd}
\end{equation}
being $E_{i}$ the single-particle energy of neutron $i$, $\mu_i$ its chemical potential and $T$ the temperature of the system. The single-particle energy $E_{i}$ and the chemical potential $\mu_i$ should be obtained from a particular model of neutron matter. In this work, as it has been already said,
to describe the bulk and single-particle properties of neutron matter under the presence of a strong magnetic field we use the BHF approximation of the Brueckner--Bethe--Goldstone (BBG) non-relativistic many-body theory of nuclear matter. A detail discussion of the BHF approach can be found in~\cite{Ag14}.

Let us now focuss on the evaluation of the matrix ${\cal M}_{\nu' n',\nu n}$. Here we show the main steps on
the derivation, and we refer the interested reader to appendix \ref{MollerMatrix} for specific details. Our starting point is the following Lagrangian density written in terms of a current-current interaction as:
\begin{equation}
\mathcal{L}=\frac{1}{\sqrt{2}} G_F \biggl( \bar{\psi}_{\nu'} \gamma^\mu \frac{1}{2}\left(1-\gamma_5 \right)\psi_{\nu}\biggr) \biggl(\bar{\psi}_{n'} \gamma_\mu \left(C_V-C_A\gamma_5 \right) \psi_{n} \biggr)
\ .
\end{equation}
Here $G_F \simeq 1.436 \times 10^{-49}$ erg cm$^{-3}$ is the Fermi weak coupling constant and the quantities $C_{V}=-1/2$ and $C_{A}=-1.23/2$ are the vector and axial--vector couplings, respectively.
The matrix ${\cal M}_{\nu' n',\nu n}$ can be written from this Lagrangian density as:
\begin{equation}
{\cal M}_{\nu' n',\nu n}= \frac{1}{\sqrt{2}} G_F \biggl(\overline{u}_{\nu'}\gamma^\mu \frac{1}{2}\left(1-\gamma_5 \right) u_{\nu}\biggr) \biggl(\overline{u}_{n'}\gamma_\mu \left(C_V-C_A\gamma_5 \right)u_{n}\biggr) \ .
\label{matrixM}
\end{equation}
It is convenient to express the square of this matrix as the contraction of a leptonic ($l^{\mu\alpha}$) and an hadronic ($H_{\mu\alpha}$) two-rank tensor,
\begin{equation}
|{\cal M}_{\nu' n',\nu n}|^{2}= \frac{1}{2} G_F^{2} \, l^{\mu\alpha} H_{\mu\alpha} \, ,
\end{equation}
with
\begin{equation}
l^{\mu\alpha}=\biggl(\overline{u}_\nu \gamma^{\mu} \frac{1}{2} \left(1-\gamma_5\right)u_{\nu'} \biggr)
\biggl(\overline{u}_{\nu'}\gamma^\alpha \frac{1}{2} \left(1-\gamma_5 \right)u_{\nu}\biggr),
\label{lepton}
\end{equation}
and
\begin{equation}
H_{\mu\alpha}=\biggl(\overline{u}_{n} \left(C_V+C_A \gamma_5 \right)\gamma_\mu u_{n'}\biggr)
\biggl(\overline{u}_{n'}\gamma_\alpha \left(C_V-C_A\gamma_5 \right) u_{n}\biggr) \ .
\label{hadron}
\end{equation}
Note that in Eq.\ (\ref{hadron}) the summation over the spin quantum number is implicit. If neutron matter is not polarized then $|{\cal M}_{\nu' n',\nu n}|^{2}$ (and consequently $\sigma(p_{\nu})/V$) can be simply obtained from Eqs.\ (\ref{lepton}) and (\ref{hadron}). However, the presence of a magnetic field induces a (partial) spin polarization of the system and, therefore, in this case this summation should be split between neutrons with spin up and down. To take this into account, we employ the spin projection operator,
$\Lambda_{s}=\frac{1}{2} \left(1+ \gamma_5\slashed{w}_s\right)$, with
the four-vector $w_s=\left(0,0,0,s \right)$, where $s=+1$ ($-1$) projects into the spin up (down) configuration.
Using this operator in Eq.~(\ref{hadron}), we have,
\begin{equation}
H_{\mu\alpha}^{s}=\biggl(\overline{u}_{n} \frac{1}{2} \left(1+\gamma_5 \slashed{w}_s\right)\left(C_V+C_A \gamma_5 \right)\gamma_\mu u_{n'}\biggr)
\biggl(\overline{u}_{n'}\gamma_\alpha \left(C_V-C_A\gamma_5 \right) \frac{1}{2} \left(1+ \gamma_5\slashed{w}_s\right)u_{n}\biggr) \ .
\end{equation}
Note that the action of the operator $\Lambda_{s}$, generates the tensors $H^{-}_{\mu\alpha}$ for neutrons with spin down and $H^{+}_{\mu\alpha}$ for neutrons with spin up. The total hadronic tensor can then be written as:
\begin{equation}
H_{\mu\alpha}=\frac{\left(1-{\cal A} \right)}{2} H^{-}_{\mu\alpha} +\frac{\left(1+{\cal A} \right)}{2} H^{+}_{\mu\alpha} \ ,
\label{had2}
\end{equation}
where the tensors $H^{-}_{\mu\alpha}$ and $H^{+}_{\mu\alpha}$ are weighted according to the degree of polarization of the system given by the spin asymmetry defined as,
\begin{equation}
{\cal A} =\frac{\rho_{+}-\rho_{-}}{\rho_{+}+\rho_{-}} \ ,
\label{assim}
\end{equation}
with $\rho_{+}$ ($\rho_{-}$) being the density of neutrons with spin up (down). Note that the value ${\cal A}=0$ corresponds to unpolarized neutron matter, whereas ${\cal A}=+1$  or ${\cal A}=-1$ means that the system is in a completely polarized state with all the spins up or down, respectively. Partially polarized states correspond to values of ${\cal A}$ between $-1$ and $+1$.

Contracting now the hadronic tensor of Eq.\ (\ref{had2}) with the leptonic one, we obtain,
\begin{equation}
|{\cal M}_{\nu' n',\nu n}|^2 =  |{\cal M}_{\nu' n',\nu n}^-|^2 +
|{\cal M}_{\nu' n',\nu n}^+|^2 \ ,
\end{equation}
where
\begin{eqnarray}
|\mathcal{M}_{\nu'n',\nu n}^+|^2&=&16 G_F^2 \,
\frac{1+{\cal A}}{2} \,\biggl( C_V^2 \biggl(
\left(p_{\nu'}\cdot p_{n'} \right) \left(p_{\nu} \cdot p_{n}
\right)+ \left(p_{\nu'} \cdot p_{n} \right)\left(p_\nu\cdot
p_{n'}\right)-\left(p_{\nu'}\cdot p_{\nu}\right)
(m^*_+)^2\nonumber\\ &&-m^*_+ \left(p_{\nu z}
\left(p_{\nu'}\cdot\left(p_{n'} -p_n\right)\right)
-p_{\nu'z}\left(p_{\nu}\cdot\left(p_{n'}-p_{n}\right)\right)\right)\biggr)\nonumber\\
&&+C^2_A \biggl( \left(p_{\nu'} \cdot
p_{n'}\right)\left(p_{\nu}\cdot p_n\right)+ \left(p_{\nu'} \cdot
p_n \right)\left(p_\nu \cdot p_{n'}\right)+\left(p_{\nu'} \cdot
p_\nu \right) (m^*_+)^2 \nonumber \\ &&-m^*_+ \left(p_{\nu
z}\left(p_{\nu'}\cdot\left(p_{n'}
+p_{n}\right)\right)-p_{\nu'z}\left(p_{\nu}\cdot\left(p_{n'}+p_n\right)\right)
\right)\biggr)\nonumber\\ &&-2 m^*_+ C_V C_A
\left(\left(p_{n'}\cdot p_{\nu'}\right) p_{\nu z}+\left(p_n
\cdot p_\nu \right) p_{\nu' z} \right)\biggr)
\label{mreld}
\end{eqnarray}
and
\begin{eqnarray}
|\mathcal{M}_{\nu'n',\nu n}^-|^2&=&16 G_F^2 \,
\frac{1-{\cal A}}{2}
\,\biggl(C_V^2\biggl(\left(p_{\nu'}\cdot
p_{n'}\right)\left(p_\nu\cdot p_n\right)+\left(p_{\nu'}\cdot
p_n\right)\left(p_\nu \cdot p_{n'}\right)-\left(p_{\nu'} \cdot
p_\nu\right) (m^*_-)^2 \nonumber\\ &+& m^*_-\left(p_{\nu
z}\left(p_{\nu'}\cdot\left(p_{n'}-p_n\right)\right) -p_{\nu'
z}\left(p_\nu\cdot\left(p_{n'}-p_n\right)\right)\right)\biggr)\nonumber\\
&+&C^2_A\biggl(\left(p_{\nu'}\cdot
p_{n'}\right)\left(p_{\nu}\cdot p_n\right)+\left(p_{\nu'}\cdot
p_n\right)\left(p_\nu \cdot p_{n'}\right)+\left(p_{\nu'}\cdot
p_\nu\right)(m^*_-)^2 \nonumber\\ &+& m^*_-\left(p_{\nu
z}\left(p_{\nu'}\cdot\left(p_{n'} +p_{n}\right)\right) -p_{\nu'
z}\left(p_\nu
\cdot\left(p_{n'}+p_n\right)\right)\right)\biggr)\nonumber\\
&+&2 m^*_- C_V C_A\left(\left(p_{n'}\cdot p_{\nu'}\right)p_{\nu
z}+\left(p_{n'}\cdot p_\nu \right)p_{\nu'z}\right) \biggr) \ ,
\label{mrelu}
\end{eqnarray}
being $m^{*}_{+}$ and $m^{*}_{-}$ the effective mass of neutrons with spin up and down, respectively (see Eq.\ (\ref{eq:em})).

\begin{figure}[t]
\begin{center}
    \includegraphics[width = 0.57 \textwidth]{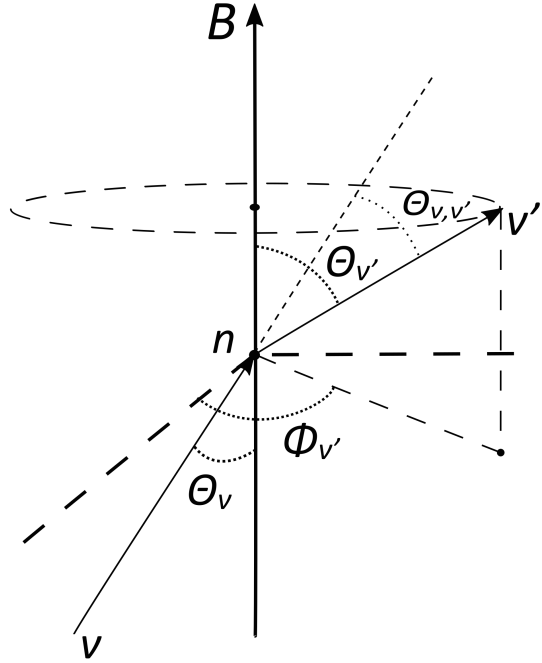}
\vskip 2mm
\caption{Geometry of the scattering process. The magnetic field defines the $z$--axis. The incoming neutrino, $\nu$, has
polar angle $\theta_{\nu}$ and without loss of generality we take its azimuthal angle $\phi_{\nu}$ equal to zero.
For the outgoing neutrino $\nu'$, we have a polar angle $\theta_{\nu'}$ and an azimuthal angle $\phi_{\nu'}$. The angle between
$\nu$ and $\nu'$ is $\theta_{\nu\nu'}$ defined through Eq.\ (\ref{eq:cosnunup}).
Note that we have neglected the neutron $n$, momenta.}
\label{figme2}
\end{center}
\end{figure}

These expressions are fully relativistic. However, we are using a non-relativistic many-body approach to describe the single-particle and bulk properties of neutron matter, therefore, to be consistent we should take the non--relativistic limit of these expressions.
Choosing the $z$--axis along the direction of the magnetic field, this limit can be obtained by using the following relations:
\begin{eqnarray}
\left(p_{n} \cdot p_{n'} \right)& \cong & (m^{*}_{\pm})^2\nonumber\\
\left(p_{n} \cdot p_{\nu} \right)& \cong & m^{*}_{\pm} E_{\nu}\nonumber\\
\left( p_{\nu'} \cdot p_{n'}\right)& \cong &m^{*}_{\pm} E_{\nu'}\nonumber\\
\left(p_{\nu} \cdot p_{\nu'} \right)& = & E_{\nu} E_{\nu'} \left(1-\cos \theta_{\nu\nu'} \right)\nonumber\\
p_{\nu z}& = & E_{\nu} \cos \theta_{\nu} \nonumber\\
p_{\nu' z} & =& E_{\nu'} \cos \theta_{\nu'} \ ,
\end{eqnarray}
where  $\theta_{\nu}$ ($\theta_{\nu'}$) is the angle between the incoming (outgoing) neutrino with the magnetic field and $\theta_{\nu\nu'}$ is the angle between the direction of the incoming and the outgoing neutrino. We note that, in the above relations, $m^*_+$ is used when evaluating the non--relativistic limit of Eq.\ (\ref{mreld}), and $m^*_-$ when taking that of Eq.\ (\ref{mrelu}). We note also that the neutron momenta are neglected when evaluating the matrices ${\cal M}^{\pm}_{\nu' n',\nu n}$. The geometry of the scattering process is shown in Fig.~\ref{figme2} . The non--relativistic limits of Eqs.\ (\ref{mreld}) and (\ref{mrelu}) then read,
\begin{eqnarray}
|{\cal M}_{\nu' n',\nu n}^+|^2&=&16 \, G_F^2 \,
\frac{\left(1+{\cal A} \right)}{2} \, (m^{*}_{+})^2 E_{\nu} E_{\nu'}
\Biggl( \left(C_V^2+3C_A^2 \right)+\left(C_V^2-C_A^2\right)
\cos \theta_{\nu\nu'}  \nonumber\\ &+& 2C_A\biggl( \left(C_A +C_V \right)
\cos \theta_{\nu} +\left(C_V-C_A \right) \cos \theta_{\nu'} \biggr) \Biggr)
\label{mnrelp}
\end{eqnarray}
and
\begin{eqnarray}
|{\cal M}_{\nu' n',\nu n}^-|^2&=&16 \, G_F^2 \,
\frac{\left(1-{\cal A} \right)}{2} \, (m^{*}_{-})^2 E_{\nu} E_{\nu'}
\Biggl( \left(C_V^2+3C_A^2 \right)+\left(C_V^2-C_A^2\right)
\cos \theta_{\nu, \nu'}  \nonumber\\ &-& 2C_A\biggl( \left(C_A +C_V \right)
\cos \theta_{\nu} +\left(C_V-C_A \right) \cos \theta_{\nu'} \biggr) \Biggr) \ .
\label{mnrelm}
\end{eqnarray}
Note that by construction, these expressions do not depend on the momentum of the incoming and outgoing neutron, since as mentioned before they were neglected in their derivation. Similar expressions can be found in other works (see, {\it e.g.,} Refs.\ \cite{Ar99,An03}).

Finally, the total cross section per unit volume is given by the sum of two contributions:
\begin{equation}
\frac{\sigma(p_{\nu})}{V} = \frac{\sigma^{+}(p_{\nu})}{V}+\frac{\sigma^{-}(p_{\nu})}{V} \ ,
\label{eq:sumcont}
\end{equation}
where each one of them, $\sigma^{\pm}(p_{\nu})/V$, is simply obtained by replacing
Eqs.\ (\ref{mnrelp}) and (\ref{mnrelm}), into Eq.\ (\ref{cross1}) reading,
\begin{eqnarray}
\frac{\sigma^{\pm}(p_{\nu})}{V} & = & G_F^2 \, \frac{\left(1 \pm {\cal A} \right)}{2} \,
\int \frac{d {\vec p_{\nu'}}}{(2 \pi)^{3}} \biggl( \left(C_V^2+3C_A^2 \right)+\left(C_V^2-C_A^2\right)
\cos \theta_{\nu\nu'}
\nonumber \\
&\pm& 2C_A\biggl( \left(C_A +C_V \right)
\cos \theta_{\nu} +\left(C_V-C_A \right) \cos \theta_{\nu'} \biggr) \biggr) \,
{\cal S}^{0}_{\pm}(q_{0},{\vec q},T)\ .
\label{crossf}
\end{eqnarray}
Here we have used the delta function $\delta^{(3)}(\vec p_{\nu}+\vec p_{n}-\vec p_{\nu'}-\vec p_{n'})$ to integrate over the momentum $\vec p_{n'}$ of the outgoing neutron. ${\cal S}^{0}_{\pm}(q_{0},{\vec q},T)$ is the structure function describing the response of neutron matter to the excitations induced by neutrinos which reads:
\begin{equation}
{\cal S}^{0}_{\pm}(q_{0},{\vec q},T) = \frac{1}{(2 \pi)^{2}} \int d {\vec p_{n}}
f_{n}^\pm({\vec p_{n}}, T) (1-f_{n'}^\pm({\vec p_{n}+{\vec q}}, T))
\delta(q_{0}+E_{n}^\pm({\vec p_{n},T})-E_{n'}^\pm({\vec p_{n}+{\vec q}},T)) \ ,
\label{strufun}
\end{equation}
being $q_{0}=E_{\nu} - E_{\nu'}$ and ${\vec q}={\vec p}_{\nu}-{\vec p}_{\nu'}$. Note that, for clarity, in the above expression we have explicitly indicated the spin projection of the neutron in the distribution functions and the single--particle energies.
An analytical expression of ${\cal S}^{0}_{\pm}(q_{0},{\vec q},T)$ can be obtained if the momentum dependence of the neutron single-particle energies is quadratic. Although this is not the case of the BHF approach, when calculating the structure function we approximate the neutron single-particle energy $E^\pm({\vec p},T)$ by the quadratic function,
\begin{equation}
E^\pm(\vec p,T)\approx\frac{|\vec p|^2}{2m_\pm^*}+U^\pm(\vec p=\vec 0,T) \, ,
\label{eq:paral}
\end{equation}
where $U^\pm(\vec p=\vec 0,T)$ is the BHF single-particle potential, which represents the average potential ``felt" by a neutron with spin projection $s=\pm 1$ in the nuclear medium (see {\it e.g.} Eq.~(8) in~\cite{Ag14}), evaluated at zero momentum  and
\begin{equation}
\frac{m_\pm^*}{m}=\frac{|\vec p|}{m}\left(\frac{dE^\pm(\vec p,T)}{dp}\right)^{-1}\Big|_{|\vec p|=p_{F_\pm}} \, ,
\label{eq:em}
\end{equation}
is the effective mass of neutrons with spin up or down, being $m$ the neutron bare mass and $p_{F_\pm}$ is the Fermi momentum of a neutron with spin projection $\pm$. Assuming this quadratic dependence of the neutron single-particle energies, the analytic expression of the structure function reads (see {\it e.g.} Refs.\ \cite{ReddyT,Na99,Ma03}):
\begin{equation}
{\cal S}^{0}_{\pm}(q_{0},{\vec q},T)=\frac{1}{\pi}\frac{1}{1-e^{-q_0/T}}\frac{(m^*_\pm)^2T}{4\pi q}
\mbox{ln}\left(\frac{1+e^{(A_\pm+q_0/2)/T}}{1+e^{(A_\pm-q_0/2)/T}} \right) \ ,
\label{eq:sfparal}
\end{equation}
where $A_\pm=\mu_\pm-m^*_\pm q_0^2/2q^2-q^2/8m^*_\pm$.

Before we discuss our numerical results for the neutrino mean free path $\lambda$, it is worth to
make some general considerations on Eq.\ (\ref{crossf}). Let us consider first the
non--polarized case (${\cal A}=0$). Without polarization we have
${\cal S}^{0}_{-}={\cal S}^{0}_{+}={\cal S}^{0}$, since in this case the single-particle energy of a neutron
is independent of its spin orientation. From Eq.~(\ref{crossf}), is then easy to obtain,
\begin{eqnarray}
\frac{\sigma(p_{\nu}) }{V}}\Big| _{{\cal A}=0 & = & G_F^2 \,
\int \frac{d {\vec p_{\nu'}}}{(2 \pi)^{3}} \biggl( C_V^2 (1+\cos \theta_{\nu\nu'} ) +C_A^2(3-\cos \theta_{\nu\nu'} )\biggr)
 {\cal S}^{0}(q_{0},{\vec q},T)
\label{crossnp}
\end{eqnarray}
which is the expression frequently found in the literature. Comparing this expression with Eq.\ (\ref{crossf})
we see that the new terms due
to the neutron polarization are the ones proportional to $\cos \theta_{\nu}$ and $\cos \theta_{\nu'}$. We note that since the integration is done over ${\vec p}_{\nu'}$, the contribution to the cross section from the term proportional to $\cos \theta_{\nu'}$ is
almost negligible. Even though is not zero, since ${\cal S}^{0}_{\pm}$ itself depends implicitly on $\cos \theta_{\nu'}$ through the transfer momentum $\vec q$ which involves the angle $\theta_{\nu\nu'}$, whose
cosine can be easily written as (see Fig.\ \ref{figme2}),
\begin{equation}
\cos \theta_{\nu\nu'} = \sin \theta_{\nu} \sin \theta_{\nu'}  \cos \phi_{\nu'} +\cos \theta_{\nu} \cos \theta_{\nu'} \ .
\label{eq:cosnunup}
\end{equation}
A final obvious comment, is that the cross section depends on the energy and momentum of the incoming
neutrino. Note, in particular, that if the momentum of the incoming neutrino is perpendicular to the magnetic field then $\cos \theta_{\nu}=0$ and one expects no appreciable differences
with respect to the unpolarized case.

\section{Results and discussion}
\label{Results}

In the following we present results for the mean free path of neutrinos in homogeneous hot neutron matter under the presence of strong magnetic fields. Results are shown for densities in the range $0.05 \leq\rho \leq 0.4$ fm$^{-3}$ corresponding approximately to the outer core region a neutron star, several temperatures up to T=30 MeV, and three values of the magnetic field intensity B=0, $10^{18}$ and $2.5\times 10^{18}$ G. As we have already mentioned, our description of the bulk and single-particle properties of hot and magnetized neutron matter is based on the non-relativistic BHF approach developed in~\cite{Ag14} using, in particular, the Argonne V18 nucleon-nucleon potential \cite{argonne} supplemented with the Urbana IX three-nucleon force \cite{urbana}.

\begin{figure}[t]
\begin{center}
 \vskip -10mm
    \includegraphics[scale=0.47]{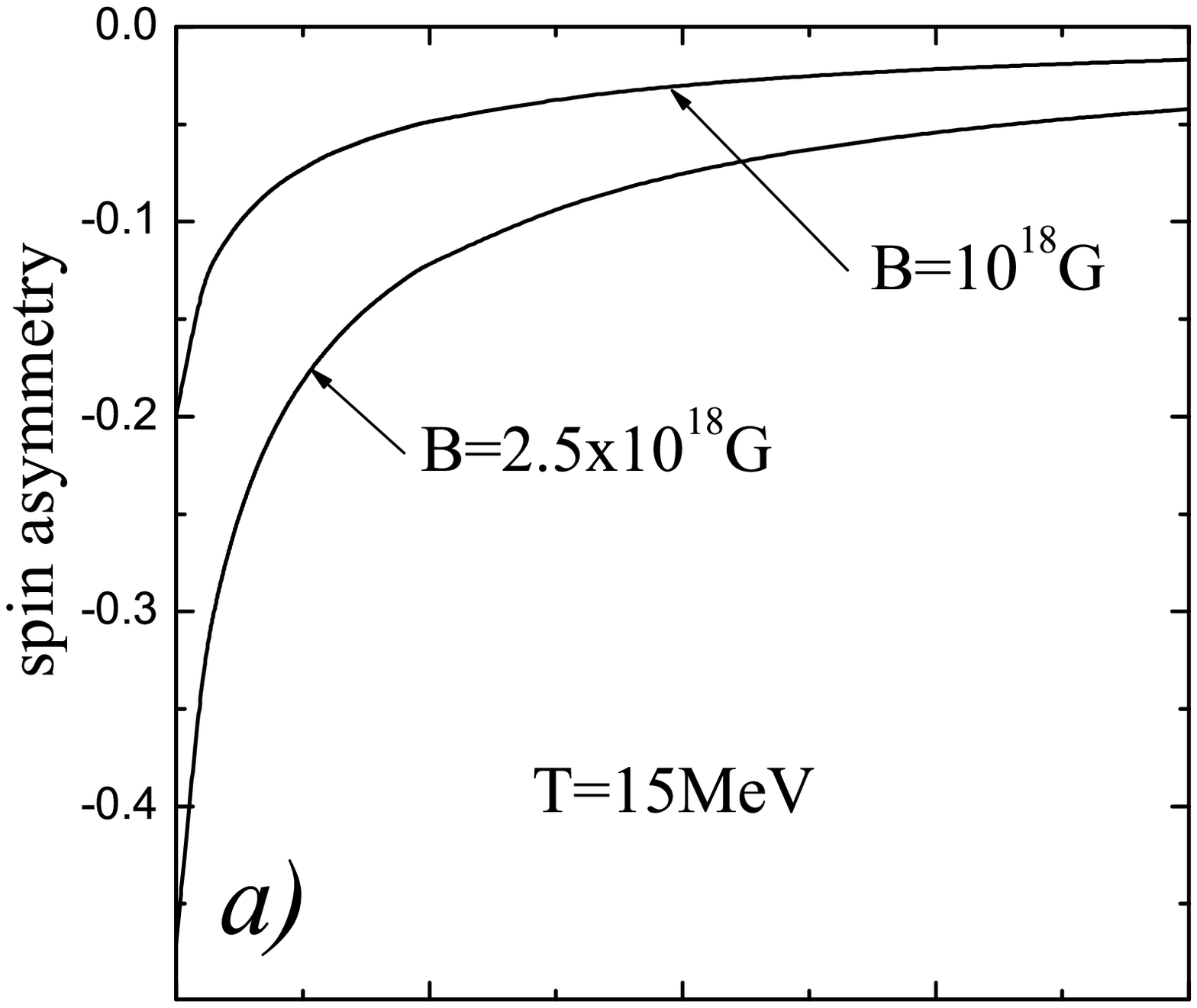}
\vskip -6cm
    \includegraphics[scale=0.47]{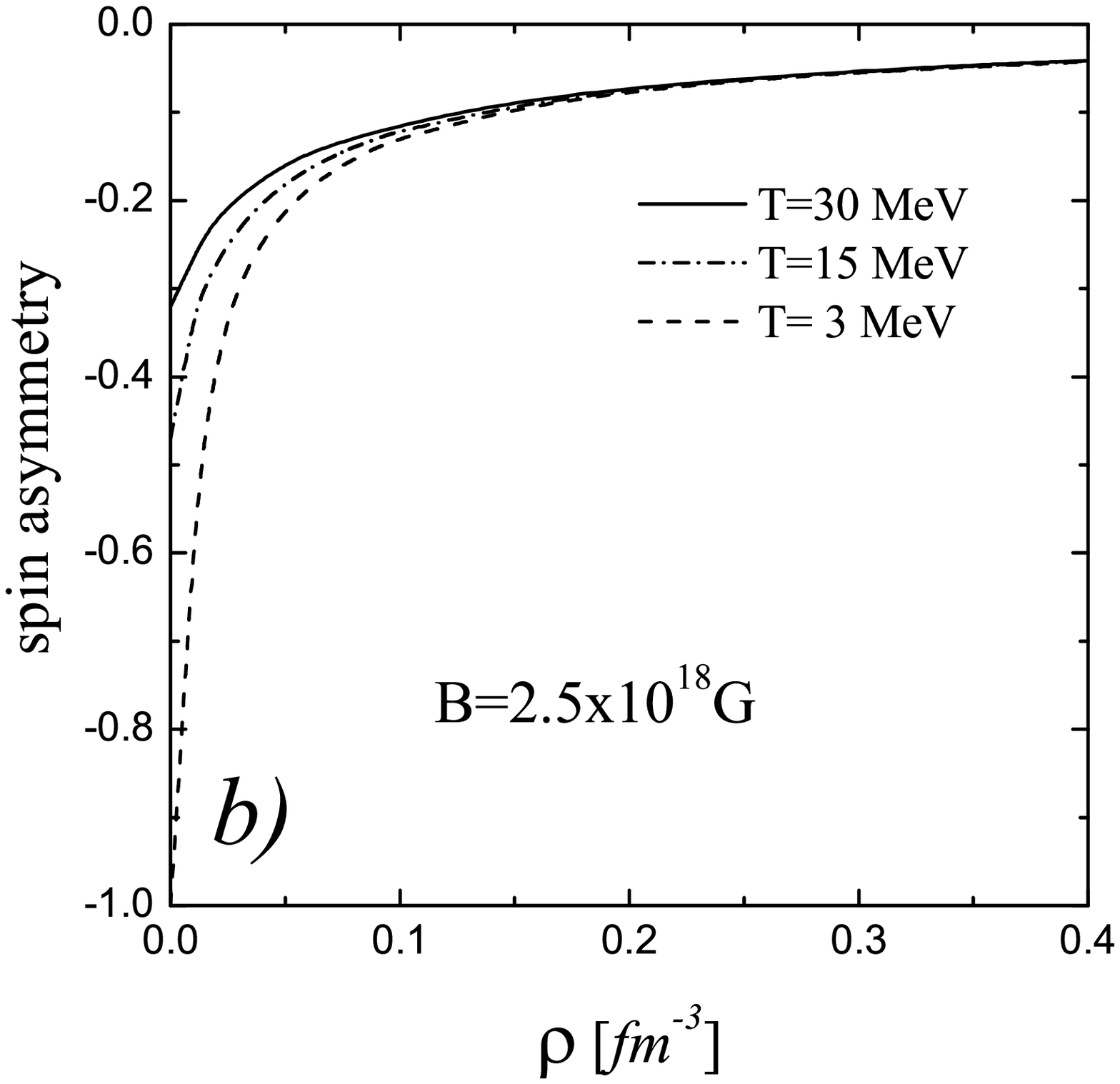}
\vskip 2mm
\caption{Density dependence of the spin asymmetry ${\cal A}$ for: a) different values of the magnetic field strength and b) different values of the temperature. }
\label{figme3}
\end{center}
\end{figure}

Before discussing our results for the neutrino mean free path, we analyze first the spin asymmetry ${\cal A}$ of the system, the effective masses of neutrons with spin up and down, and the structure function ${\cal S}^{0}_{\pm}(q_0,\vec q,T)$ predicted by our BHF model for different temperatures and magnetic field intensities.
As it was mentioned in the previous section, the spin asymmetry ${\cal A}$ characterizes the degree of polarization of the system. The physical state is obtained by minimizing the Helmhotz free energy density of the system with respect to ${\cal A}$
for fixed values of the density, the temperature and the magnetic field. We note that this minimization implies that in the physical state the chemical potential of neutrons with spin up and spin down is the same, {\it i.e.,} there is only one chemical potential which is associated to the conservation of total baryonic number. We note also that the degree of polarization of the physical state of the system is the result of the competition between the strong interaction that, together with the temperature, favor the non-polarized state as the physical one, and the magnetic field that tries to align all the spins antiparallel to it. In Fig.\ \ref{figme3} we show the spin asymmetry corresponding to the physical state of the system as a function of density for several temperatures and two values of the magnetic field strength.
Although it is not shown in the figure, in the absence of a magnetic field the physical state of the system corresponds to the non-polarized case (${\cal A}=0$) for all densities and temperatures.  For low densities and temperatures, one expects that the system would be completely polarized (${\cal A}=-1$) up to a given density, above which it becomes partially polarized with a predominance of spin-down states ($-1<{\cal A}<0$). Within our range of temperatures, ${\cal A}$ grows monotonously and the system would reach the non-polarized state (${\cal A}=0$) asymptotically at high densities. A comparison of the results for B=$10^{18}$ G and B=$2.5\times 10^{18}$ G (see panel (a)) shows that the density at which the system changes from completely to partially polarized increases with B as one naively would guess. As it is seen in the panel (b) of the figure, the increase of temperature makes the system to be less polarized as one intuitively expects since it favors the disorder of the spins.

\begin{figure}[t]
\begin{center}
    \includegraphics[scale=0.53]{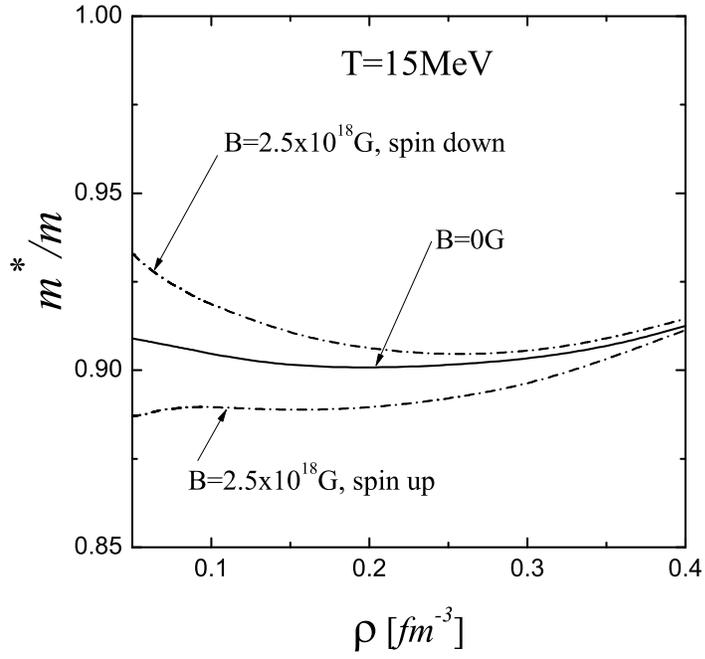}
\vskip 2mm
\caption{Density dependence of the spin up and spin down effective masses at T=15 MeV for B=0 G and B=$2.5\times 10^{18}$ G.}
\label{figme4}
\end{center}
\end{figure}

We examine in the following the neutron effective mass, which is a representative single-particle property. Although in the BHF approach the effective mass has a momentum dependence, in this work and according to the definition given in Eq.\ (\ref{eq:em}), we analyze it at the value of the Fermi momentum $p_{F_\pm}$, of a neutron with spin up or down projection. The density dependence of the effective mass of neutrons with spin up ($m^*_+$) and spin down ($m^*_-$) is shown in Fig.\ \ref{figme4}, for a temperature T=15 MeV and two values of the magnetic field strength, B=0 G and B=$2.5\times 10^{18}$ G. As it is expected, in the absence of a magnetic field we have $m^*_+=m^*_-$. Note that the magnetic field induces a splitting between $m^*_+$ and $m^*_-$ with $m^*_+<m^*_-$ over all the density range. This splitting is a direct consequence of the spin polarization dependence of the neutron single-particle potential originated by the presence of the field. The magnetic field polarizes partially the system with a spin asymmetry $-1<{\cal A}<0$ making the single-particle potential for neutrons with spin down (the most abundant component) less attractive that the one for neutrons with spin up. As shown in~\cite{Vi02b} (see, in particular, Eqs. (23) and (24) of this reference), this is due to: (i) the change in the number of pairs which a neutron with momentum $k$ and spin projection $s$ can form with the other neutrons of the system as neutron matter is polarized, and (ii) to the spin dependence of the neutron-neutron $G$-matrix in the spin polarized medium. Indeed, as the spin asymmetry decreases (becomes more negative) the single-particle potential of a spin down neutron is built from a larger number of down-down pairs that form a spin triplet state $S=1$ and, due to the Pauli principle, can only interact through odd angular momentum partial waves. Conversely, the potential of the less abundant component is built from a relatively larger number of up-down pairs which can interact both in the spin 0 and spin 1 channels. Thus, the potential of the less abundant component receives also contributions from some important attractive channels as, {\it e.g.,} the $^1$S$_0$. Finally, we would like to point out that the reason why $m^*_+<m^*_-$ can be traced back to the general issue that in a two component fermionic system the most abundant component is less correlated than the less abundant one (see {\it e.g.} \cite{hen}). In our case neutrons with spin down are more abundant than neutrons with spin up and, therefore, are expected to be less correlated. Being less correlated their effective mass should be closer to the value of the bare mass and, consequently, larger than that of neutrons with spin up, as it is in fact observed.

\begin{figure}[t]
\begin{center}
\vskip -10mm
    \includegraphics[scale=0.47]{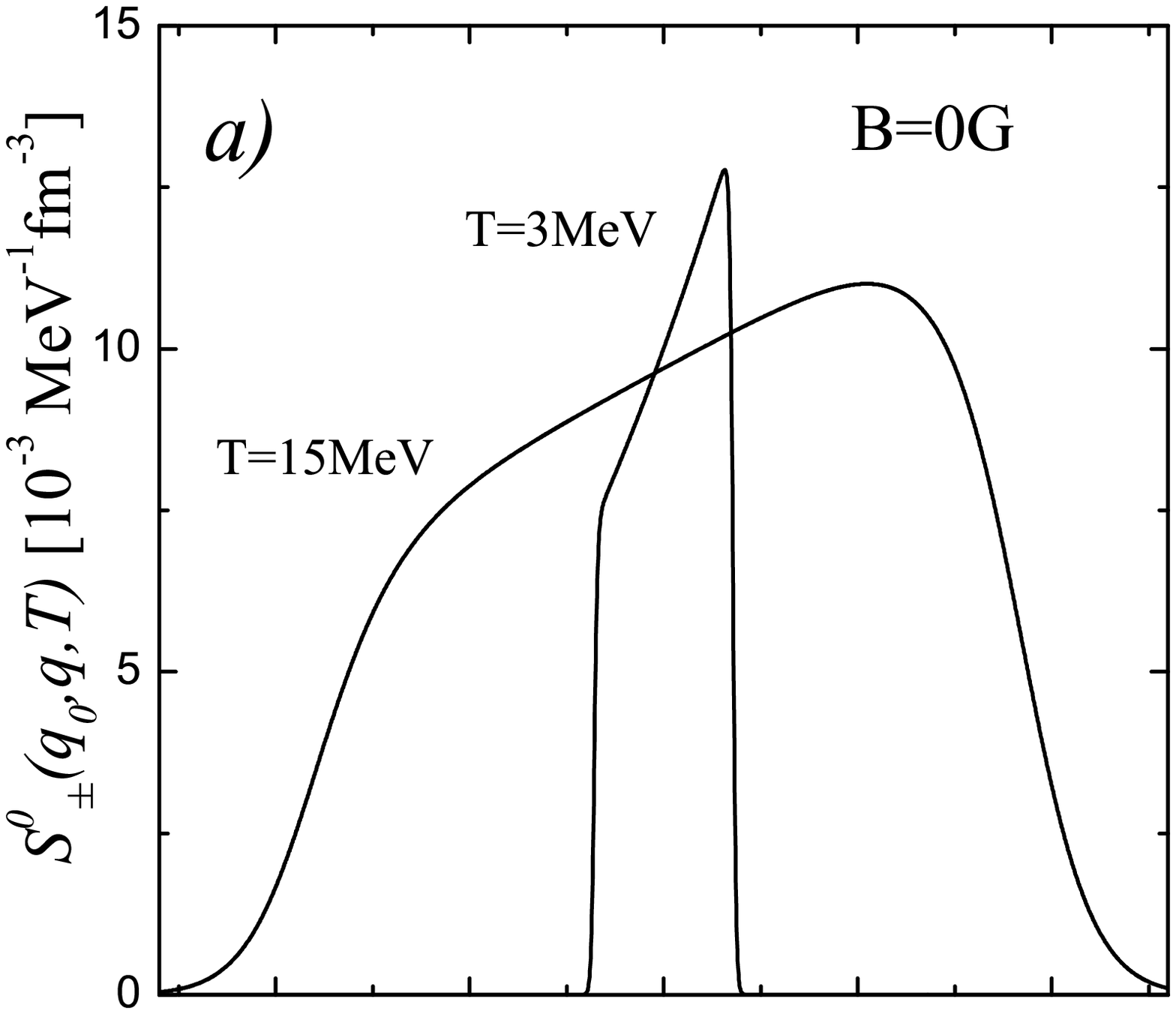}
\vskip -6cm
    \includegraphics[scale=0.47]{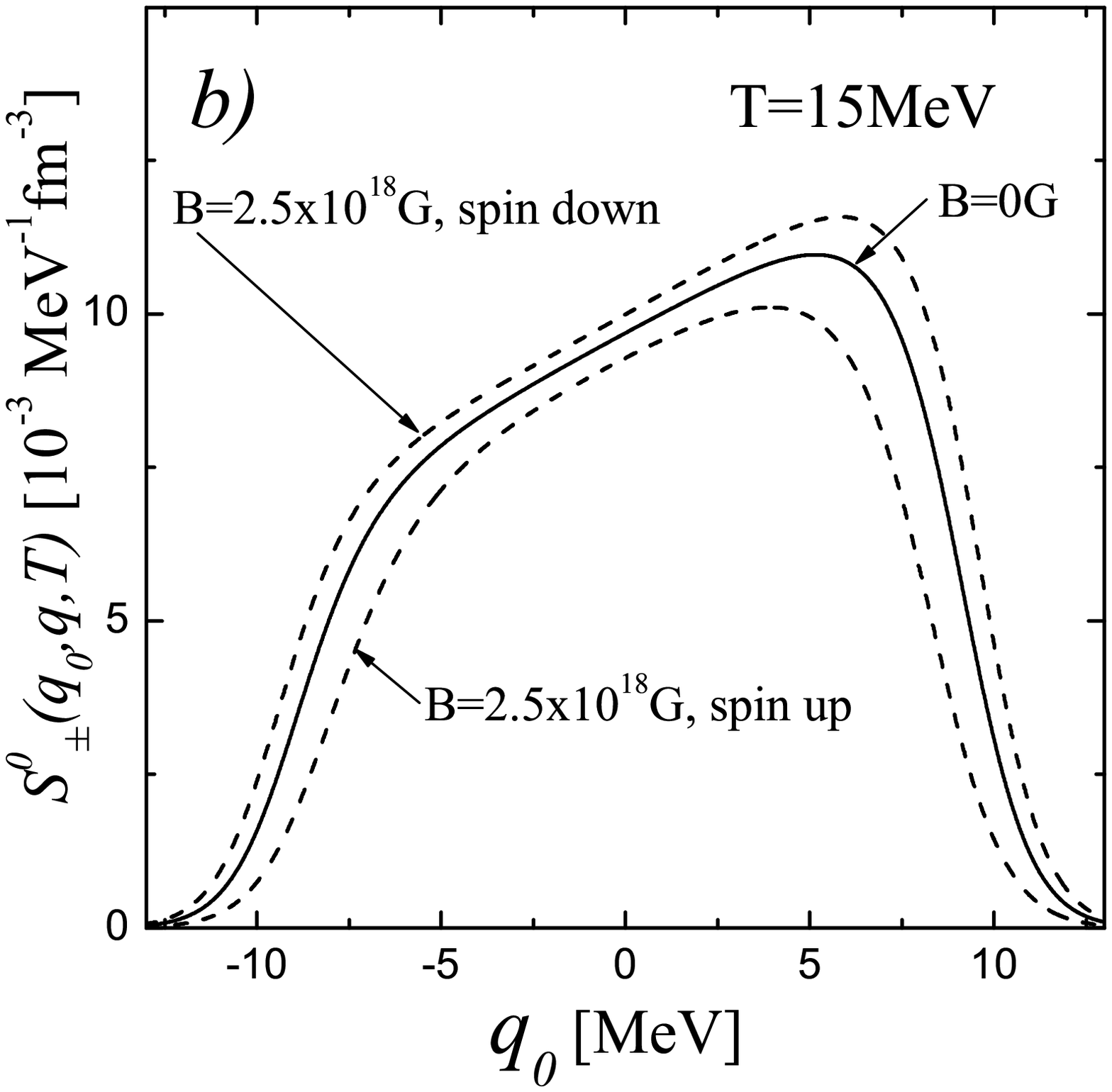}
\vskip 2mm
\caption{Energy dependence of the structure function ${\cal S}^{0}_{\pm}(q_{0},q,T)$ for $\rho=0.16$ fm$^{-3}$. Results for B=0 G with T=3 MeV and 15 MeV are shown in panel (a), whereas those for T=15 MeV and B=0 G and  B=$2.5\times 10^{18}$ G are presented in panel (b). In both panels the momentum transfer is fixed to the value $\vec q=\vec p_{\nu}/2$ with $|\vec p_{\nu}| = 3 T$.}
\label{figme5}
\end{center}
\end{figure}

Let us now give some insight into the effect of the structure function ${\cal S}^{0}_{\pm}(q_{0},{\vec q},T)$, defined in Eq.~(\ref{strufun}), on the neutrino mean free path. In Fig.\ \ref{figme5} we show ${\cal S}^{0}_{\pm}(q_{0},{\vec q},T)$ as a function of $q_0$ for a density of the system $\rho=0.16$ fm$^{-3}$. The momentum transfer is fixed to the value $\vec q=\vec p_{\nu}/2$ where the magnitude of the momentum  of the incoming neutrino $\vec p_\nu $ has been taken according to the prescription $|\vec p_{\nu}| = 3 T$, being T the temperature of the system. Results in the absence of a magnetic field for temperatures T=3 and T=15 MeV are shown in panel (a) whereas in panel (b) the structure function is shown for T=15 MeV and magnetic fields B=0 G (which will serve as a reference) and B=$2.5\times 10^{18}$ G. As it is seem in panel (a) an increase of the temperature leads to a much broader structure function with a larger area under it. The reason is simply due to the fact that the phase space of the integral in Eq.\ (\ref{strufun})
increases with temperature. Consequently, an increase of the temperature will give rise to a larger cross section and, therefore, to a smaller neutrino mean free path when integrating Eq.\ (\ref{crossf}), as we will see later. Besides the dependence of the structure function on $q_0$, $\vec q$ and $T$, from its definition (see Eq.\ (\ref{strufun})), it is clear that it depends also on the spin projection of the neutrons. This dependence leads to a splitting between ${\cal S}^{0}_{+}(q_{0},{\vec q},T)$ and ${\cal S}^{0}_{-}(q_{0},{\vec q},T)$ with ${\cal S}^{0}_{+}(q_{0},{\vec q},T)<{\cal S}^{0}_{-}(q_{0},{\vec q},T)$, as it is observed in the panel (b) of the figure. The origin of this splitting can be easily understood by looking at the analytic expression of ${\cal S}^{0}_{\pm}(q_{0},{\vec q},T)$, given in Eq.\ (\ref{eq:sfparal}), which shows that ${\cal S}^{0}_{\pm}(q_{0},{\vec q},T)$ depends quadratically on the effective mass $m^*_\pm$, but also in an implicit way though the logarithm. It can be easily shown that the explicit quadratic dependence is the dominant one and, in good approximation, one can simply assume ${\cal S}^{0}_{\pm}(q_{0},{\vec q},T)\sim (m^*_{\pm})^2F(q_{0},{\vec q},T)$ with $F(q_{0},{\vec q},T)$ encoding all the other dependencies. It is clear, therefore, that for $B\neq 0$ one has ${\cal S}^{0}_{+}(q_{0},{\vec q},T)<{\cal S}^{0}_{-}(q_{0},{\vec q},T)$ since, as we saw before, the presence of a magnetic field induces a splitting between $m^*_+$ and $m^*_-$ with $m^*_+ < m^*_-$.
In addition, when the magnetic field strength is increased, the spin asymmetry ${\cal A}$ becomes more negative (see Fig.\ \ref{figme3}) and consequently, the factors $(1+{\cal A})$ and  $(1-{\cal A})$, appearing in the expression for the spin up and down contributions to the total cross section (see Eq.\ (\ref{crossf})), decrease and increase, respectively. Therefore, an increase of the magnetic field strength will lead to a decrease of $\sigma_+$ and to an increase of $\sigma_-$ which dominates over the former giving rise, as we will show later,  to a net increase (decrease) of the total cross section
(neutrino mean free path).

\begin{figure}[t]
\begin{center}
    \includegraphics[scale=0.53]{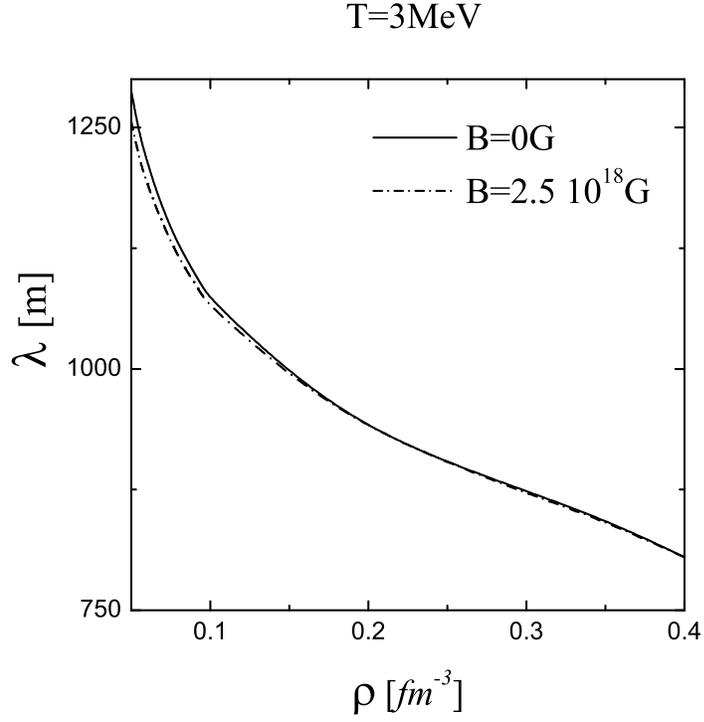}
\vskip 2mm
\caption{Neutrino mean free path as a function of the density at T=3 MeV for B=0 G and B=$2.5\times 10^{18}$G. The angle between the incoming neutrino and the magnetic field is taken at $\theta_{\nu}=\pi/2$. For the momentum of the incoming neutrino we employ $|\vec p_{\nu}| = 3 T$.}
\label{figme6}
\end{center}
\end{figure}

We will focus now our discussion on the behavior of the neutrino mean free path $\lambda$. Before starting our analysis, however, we will make a general remark. Note that in the absence of a magnetic field the total cross section (see Eq.\ (\ref{crossnp})) depends only on the energy (or the magnitude of the momentum) of the incoming neutrino but not on its direction. The reason is simply that one can always take the ${\hat z}$-axis along the direction of the outgoing neutrino to perform the integral and, therefore, the angle $\theta_{\nu\nu'}$ between the direction of both neutrinos is integrated out. This is not the case when the magnetic field is different from zero. Its presence establishes a preferred direction in the space and, consequently, in this case the total cross section (see Eqs.\ (\ref{eq:sumcont}) and (\ref{crossf})) depends both on the energy of the incoming neutrino, and on the angle $\theta_\nu$ between its momentum $\vec p_\nu$ and the direction of the magnetic field. It is interesting to note, however, that if $\vec p_\nu$ is perpendicular ({\it i.e.,} $\theta_\nu=\pi/2$) to the magnetic field then the neutrino mean free path is expected to be quite insensitive to the magnetic field. This is shown in Fig.\ \ref{figme6}, where $\lambda$ is depicted as function of the density for T= 3 MeV, $\theta_\nu=\pi/2$ and the magnetic fields B=0 G and B=$2.5\times 10^{18}$ G. As it is seem in the figure, appreciable differences are noticed only at densities below $\sim 0.15$ fm$^{-3}$. We note that for smaller magnetic field strengths no difference is observed with the B=0 G case. The reason for this low magnetic field dependence when $\theta_\nu=\pi/2$, is that the term proportional to $\cos\theta_{\nu'}$ in Eq.\ (\ref{crossf}) would cancel out except for the smooth implicit $\theta_{\nu'}$-dependence of the structure function through the angle $\theta_{\nu\nu'}$ (see Eq.\ (\ref{eq:cosnunup})) which, however, is negligible
for $\theta_\nu=\pi/2$. The only dependence on the magnetic field that remains is, therefore, that of the structure function itself which is mostly appreciable in the low/medium density region where the spin asymmetry is ${\cal A}$ larger in absolute value (see Fig.\ \ref{figme3}).

\begin{figure}[t]
\begin{center}
\vskip -30mm
    \includegraphics[scale=0.47]{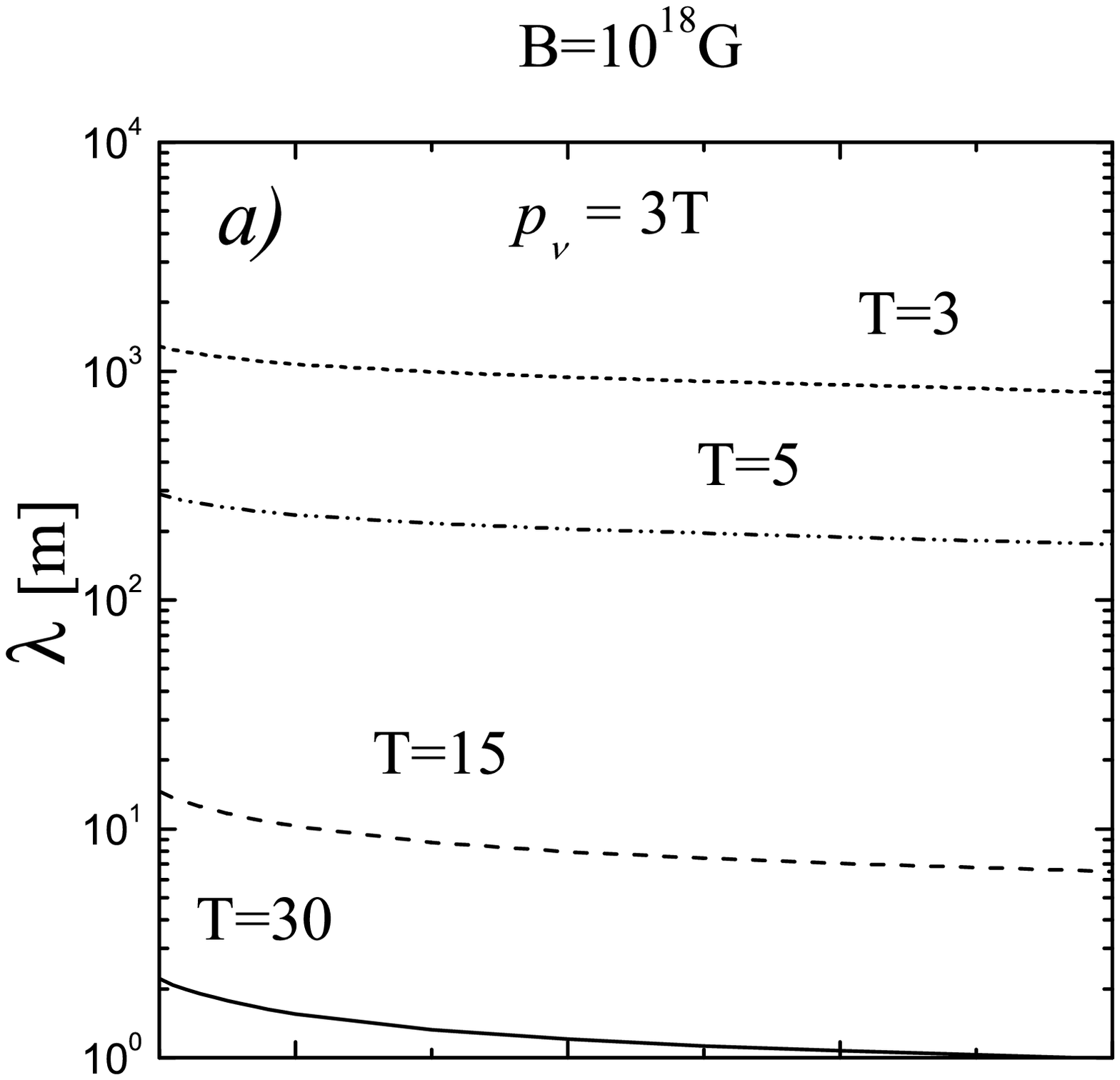}
\vskip -6cm
    \includegraphics[scale=0.47]{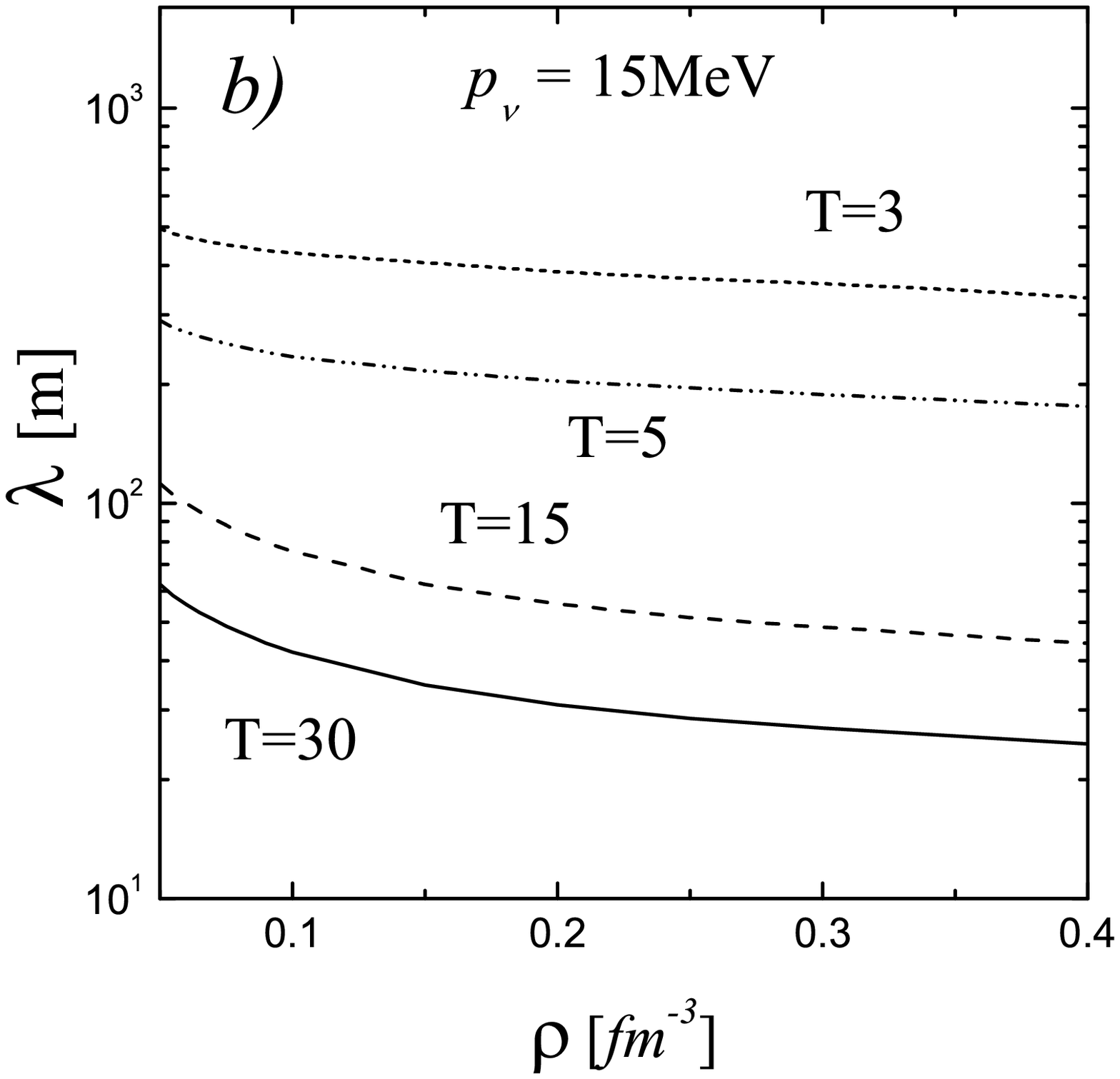}
\vskip -5mm
\caption{Neutrino mean free path as a function of the density at B=$10^{18}$ G and $\theta_{\nu}=\pi/2$ for several values of the temperature. For the momentum of the incoming neutrino we take $|\vec p_{\nu}| = 3 T$ in panel (a) and
$|\vec p_{\nu}| = 15$ MeV in panel (b).}
\label{figme7}
\end{center}
\end{figure}

We discuss now the temperature dependence of the neutrino mean free path.
In Fig.~\ref{figme7}, we show the density dependence of $\lambda$ for a magnetic field strength B=$10^{18}$ G, $\theta_\nu=\pi/2$, and the temperatures T=3, 5, 15 and 30 MeV. The momentum of the incoming neutrino is taken $|\vec p_\nu|=3T$ in panel (a) and $|\vec p_\nu|=15$ MeV in panel (b). Note that, for a fixed temperature, the larger the value of $|\vec p_\nu|$, the smaller the neutrino mean free path. This is simply due to the fact that the response of the system to the excitations induced by neutrinos, described by the structure function, is larger for larger values of the neutrino momentum. Consequently, the total cross section is larger and the neutrino mean free path smaller. As it is seen in both panels, $\lambda$ varies dramatically with temperature decreasing up to fours orders of magnitude (see panel (a)) for increasing values of the temperature. This can be easily understand from our previous analysis of the temperature dependence of the structure function ${\cal S}^{0}_{\pm}(q_{0},{\vec q},T)$. As we just saw, a larger temperature implies a larger phase space of the integral in Eq.\ (\ref{strufun}), and, therefore, a larger (smaller) total cross section (neutrino mean free path). Taking into account that the typical radius of a neutron star is of the order of 10-12 km, from these results one can easily conclude that a neutrino would unlikely interact with matter at low temperatures. In a somehow arbitrary way, we can say that from temperatures starting at T=10 MeV, one has to care of the neutrino scattering. Moreover, for T=30 MeV, multiple--scattering should be considered.

\begin{figure}[t]
\begin{center}
\vskip -10mm
    \includegraphics[scale=0.47]{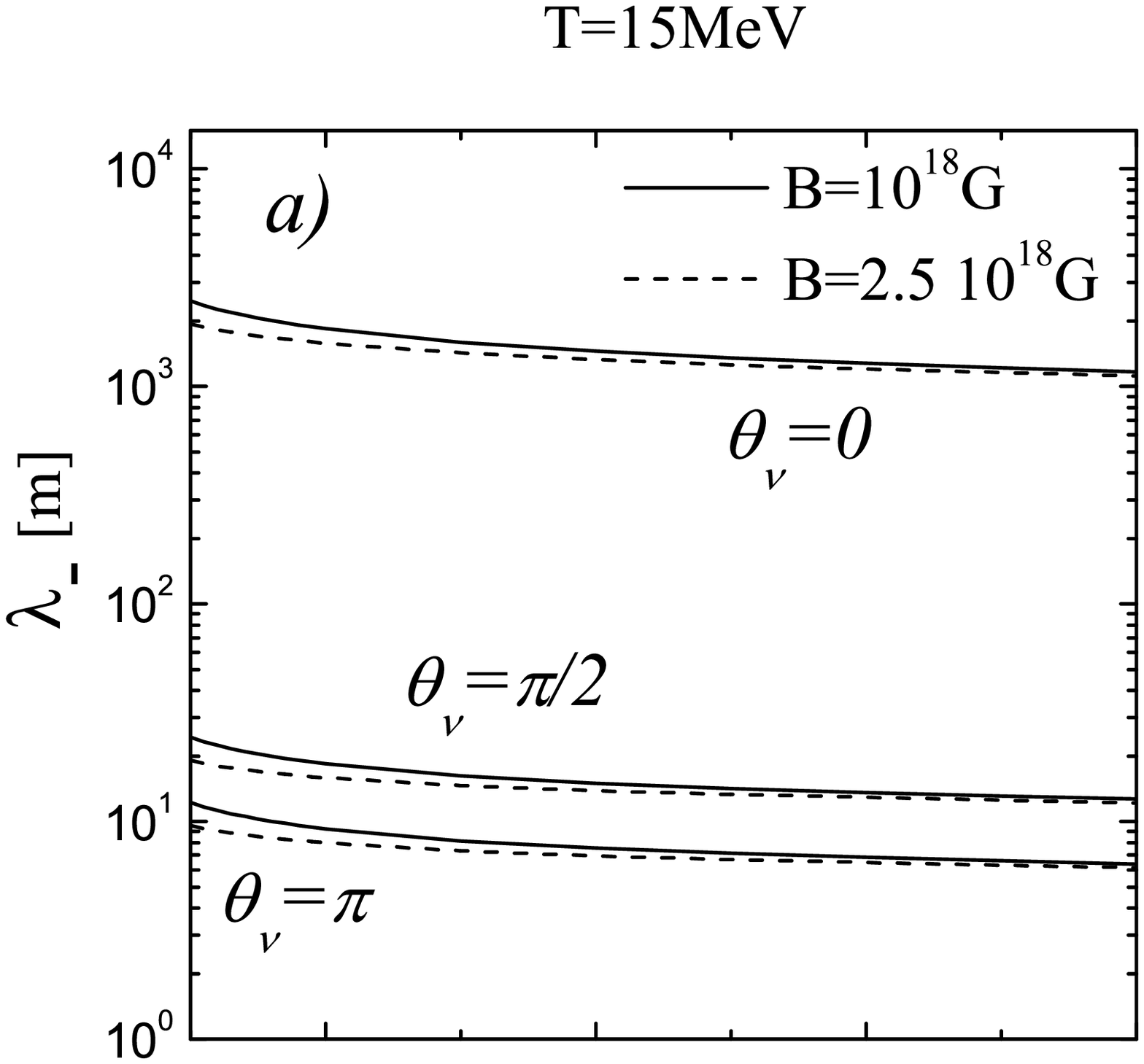}
\vskip -6cm
    \includegraphics[scale=0.47]{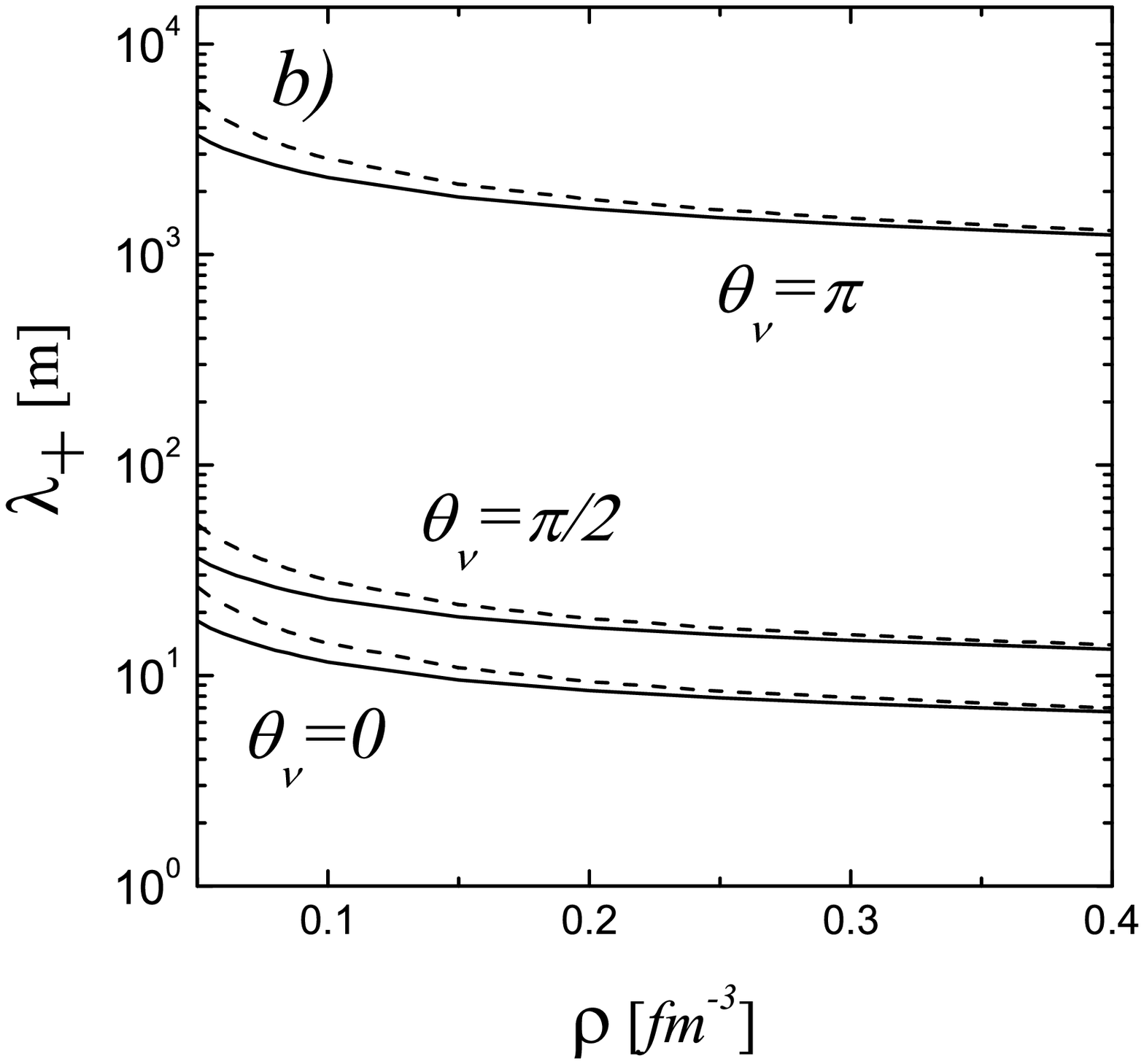}
\vskip 2mm
\caption{Neutron spin down ($\lambda_{-}$) and spin up ($\lambda_{+}$) partial contribution to the mean free path for
T=15 MeV, B=$10^{18}$G and B=$2.5\times 10^{18}$ G and different values of $\theta_{\nu}$. The momentum of the incoming neutrino is $|\vec p_{\nu}| = 3 T$.}
\label{figme8}
\end{center}
\end{figure}

We will finish this section by examining the dependence of the neutrino mean free path on the angle $\theta_\nu$. The partial contributions $\lambda_-$ (panel (a)) and $\lambda_+$ (panel (b)) to the total neutrino mean free path, due respectively to the scattering of the neutrino with a spin down or a spin up neutron and defined as $\lambda_\pm\equiv(\sigma_\pm/V)^{-1}$, are shown in Fig.\ \ref{figme8}, as a function of the density for T=15 MeV, B=0 G and $2.5\times 10^{18}$ G and the angles $\theta_\nu=0,\pi/2$ and $\pi$. We note first that both contributions vary by more than two orders of magnitude with the angle $\theta_\nu$. This huge variation cannot be understood by considering only the explicit angular factors in Eq.\ (\ref{crossf}), but it results from the combined effect of these factors and the implicit angular dependence of the structure function. Note that in polarized neutron matter, neutrons with spin down (up) are almost transparent to the neutrinos if the incoming angle of the latter is $\theta_\nu=0$ ($\pi$). Note also that $\lambda_-$ ($\lambda_+$) is shorter for $\theta_\nu=\pi$ (0).

\begin{figure}[t]
\begin{center}
\vskip -10mm
    \includegraphics[scale=0.47]{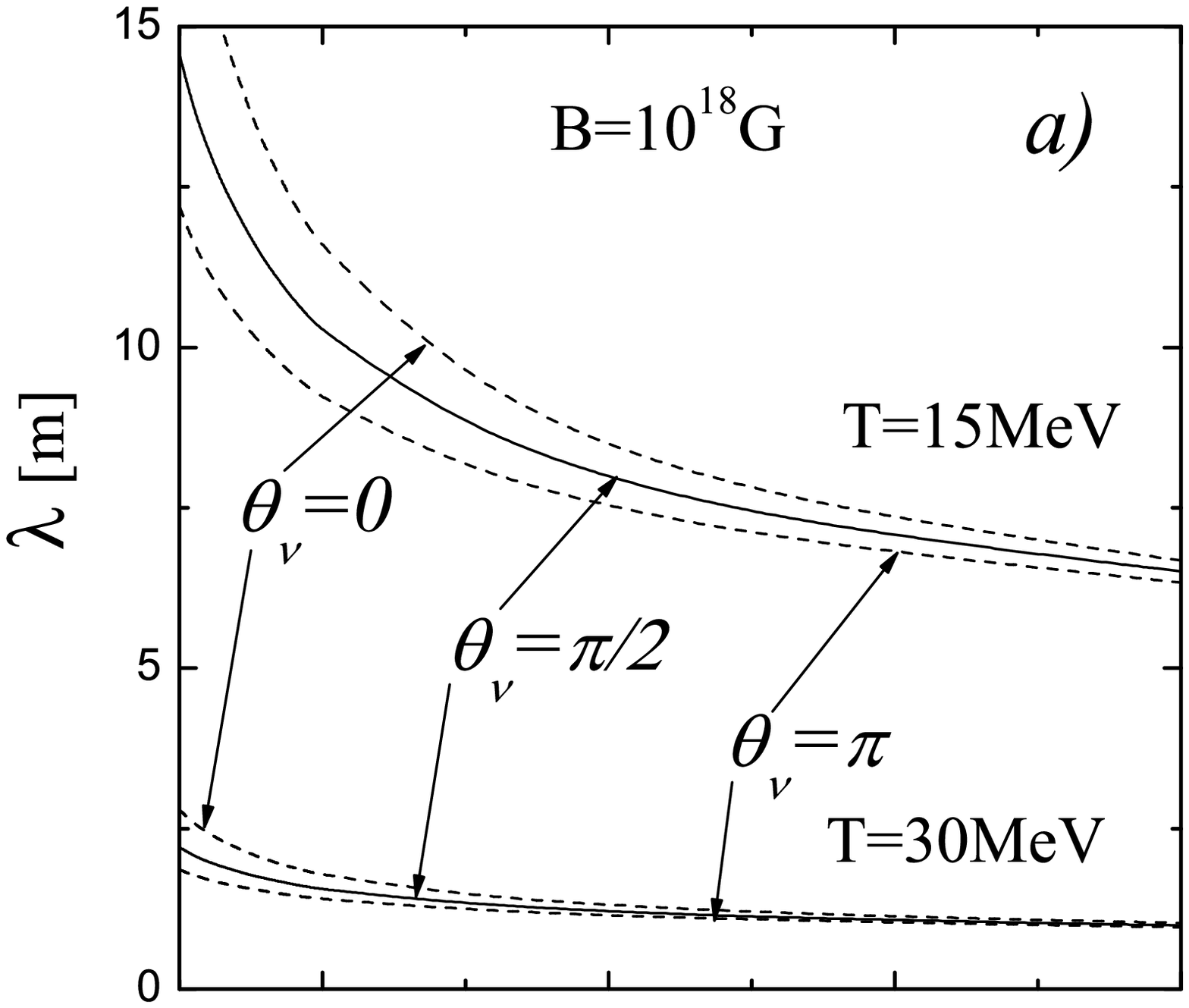}
\vskip -6cm
    \includegraphics[scale=0.47]{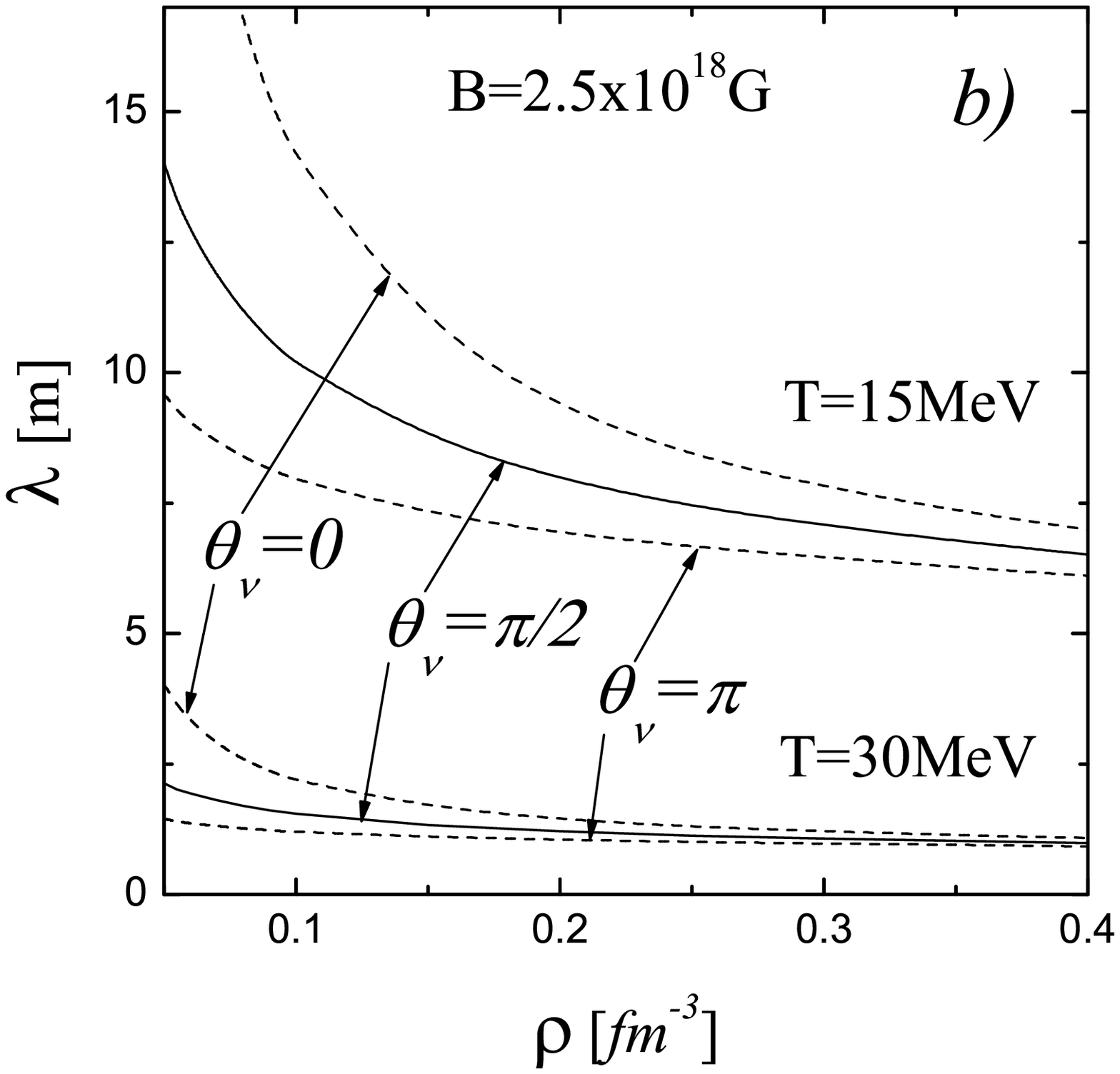}
\vskip 2mm
\caption{Dependence of the neutrino mean free path with the angle between the incoming neutrino and the magnetic field, $\theta_{\nu}$, for two values of the temperature and two values of the magnetic field intensity. The momentum of the incoming neutrino is $|\vec p_{\nu}| = 3 T$.}
\label{figme9}
\end{center}
\end{figure}

Finally, we show in Fig.~\ref{figme9}, the total mean free path for two magnetic field intensities, two temperatures and three angles. It is worth to mention, that the total mean free path can be obtained from $\lambda_-$ and $\lambda_+$ as (see Eq.\ (\ref{eq:sumcont})),
\begin{equation}
\lambda(p_\nu)=\frac{\lambda_+(p_\nu)\lambda_-(p_\nu)}{\lambda_+(p_\nu)+\lambda_-(p_\nu)} \ .
\label{eq:tnmpf}
\end{equation}

It is clear from our previous analysis that the asymmetry in the mean free path comes from the spin asymmetry factor ${\cal A}$ and the spin dependence of the structure functions ${\cal S}^{0}_{\pm}$.
Neutrinos are more transparent to polarized neutron matter when moving in a direction parallel to the magnetic field ($\theta_\nu=0$). The situation is the opposite for neutrinos that move in an anti-parallel direction ($\theta_\nu=\pi$).  In order to get a better understanding on this asymmetry in the mean free path, we define a ``mean free path asymmetry", as follows,
\begin{equation}
\chi_{\lambda} =\frac{\lambda(\theta_{\nu}=0)-\lambda(\theta_{\nu}=\pi)}{\lambda(\theta_{\nu}=\pi/2)} \ .
\label{mfpasym}
\end{equation}

\begin{table}[t]
\begin{center}
\caption{Mean free path asymmetry $\chi_{\lambda}$, as a function of the density at T=$15$MeV, for two values of the magnetic field intensity. These results are rather independent of the temperature.}
\label{asymmt}
\begin{tabular}{ccc}   \hline\hline
~~~~~~$\rho$ [fm$^{-3}$]~~~~~~ & ~~$~~~~~~\chi_{\lambda}(B=10^{18}G)~~~~~~$~~ &  $\chi_{\lambda}(B=2.5 \times 10^{18}G)$ \\ \hline
$0.050$        & $0.40$       & $1.17$                 \\
$0.100$        & $0.23$       & $0.60$               \\
$0.150$        & $0.16$       & $0.40$               \\
$0.200$        & $0.12$       & $0.29$               \\
$0.250$        & $0.09$       & $0.23$               \\
$0.400$        & $0.05$       & $0.14$               \\
\hline\hline
\end{tabular}
\end{center}
\end{table}

Note that, $\lambda(\theta_{\nu}=\pi/2)$ can be considered on practice the average value between the two extreme ones. In Table~\ref{asymmt}, we show some representative values for this ratio. Even though the asymmetry is rather small for B$=10^{18}$ G, as mentioned in the introduction, a small asymmetry in the emission of neutrinos would have a significant physical impact in a compact object. The asymmetry is more important for B$=2.5 \times 10^{18}$ G. In all cases, the asymmetry is relevant for low to medium densities. This is so, because of the dependence of the spin asymmetry parameter ${\cal A}$ and the effective masses on the density (see Figs.~\ref{figme3} and~\ref{figme4}). As the density increases, the action of the nuclear strong interaction among the neutrons overcomes the coupling of the neutrons with the magnetic field. Although this is a general behavior for all EoS--models, we should mention that the use of the Skyrme--model would lead to a bigger asymmetry for the mean free path. In this sense, our results could be interpreted as a lower limit for the discussed asymmetry.

As a final comment, we would like to mention that in a potential stellar evolution code, neutrinos would interact with neutrons either with spin up or spin down. Therefore, the partial mean free path shown in Fig.~\ref{figme8}, should be employ in the calculation. We refer to a semi--phenological model where one keeps track of an individual neutrino, using mean free path and differential cross sections evaluated with quantum mechanics. Assuming that the source of neutrinos is isotropic, the average result of many emitted neutrinos, should be consistent with the values for the total mean free path.


\section{Summary, Conclusions and Future Perspectives}
\label{Summary}
In this work we have evaluated the neutrino mean free path in neutron matter under the presence of a strong magnetic field. The description of polarized neutron matter has been done within the non--relativistic Brueckner--Hartree--Fock (BHF) approach using the Argonne V18 nucleon-nucleon potential supplemented with the Urbana IX three-nucleon force. We have considered only the neutrino scattering process in the calculation.
Starting from the Fermi Golden rule we have derived explicit expressions of the neutrino cross section per unit volume for the scattering of a neutrino with a spin up or spin down neutron.  These expressions have been obtained in the non--relativistic limit to be consistent with our description of polarized neutron matter. We have shown that
in the presence of a magnetic field the neutrino mean free path depends on the angle between the momentum $\vec p_\nu$ of the incoming neutrino and the magnetic field, leading to an asymmetry in this quantity.

In previous works by other authors, the asymmetry in the neutrino emission refers to the one originated from the differential cross section. This asymmetry and the one we have considered here are different and should be considered simultaneously to account for the actual asymmetric neutrino emission. In principle, all differential cross sections are asymmetric. However, in the absence of a preference spacial axis, the average emission from the compact object is isotropic. We have shown that this situation is altered by the presence of a magnetic field. One should be aware that the mean free path is the relevant variable in this problem: for low temperatures, the mean free path can be much larger than the size of the compact object itself. In this case, the asymmetry in the differential cross section would not be relevant, as it would be unlikely to have a collision. The total cross section (which is the inverse of the mean free path), erase the angular information of the differential cross section. That is, the asymmetry in the mean free path has a different origin than the one from the differential cross section. While the last one gives us information on the way in which the weak interaction scatters the neutrinos, the mean free path tells us about how often a neutrino interacts with a neutron.

In this analysis the temperature is the key variable. In the early stages of the cooling process of a neutron star, the temperature is high enough to ensure several collisions of the neutrinos with the neutrons before the neutrino leaves the star. It would be interesting to analyze how the asymmetry in the mean free path affects the cooling processes.
This analysis is, however, beyond the scope of the present work since among other things one should also consider the absorbtion cross section, where the neutrino is absorbed by the neutron, having a proton and an electron as the final state. The inclusion of this process is not straightforward as protons and electrons shows Landau levels in a magnetic field. We are presently working to include this mechanism.

As a final comment, we believe that the asymmetric emission of neutrinos from a magnetar has still several unexplored issues which can be relevant for the problem of the pulsar kick. In this work we have explored just one of them, namely the asymmetry in the mean free path. Apart from the absorption cross section just mentioned, another interesting point is the effect of the strong interaction over the structure function. To the best of our knowledge, this has been done only up the ring--approximation level. Our aim for the near future is to include the absorption cross section in conjunction with the asymmetry in the differential cross section to get a better understanding of the asymmetric emission of neutrinos from a magnetar.

\newpage
\appendix
\section{Evaluation of $|{\cal M}_{\nu' n',\nu n}|^2$}
\label{MollerMatrix}
In this Appendix we show some details on the evaluation of the ${\cal M}_{\nu' n',\nu n}$--matrix. We recall it expression from Eq.~(\ref{matrixM}),
\begin{equation}
{\cal M}_{\nu' n',\nu n}= \frac{1}{\sqrt{2}} G_F \biggl(\overline{u}_{\nu'}\gamma^\mu \frac{1}{2}\left(1-\gamma_5 \right) u_{\nu}\biggr) \biggl(\overline{u}_{n'}\gamma_\mu \left(C_V-C_A\gamma_5 \right)u_{n}\biggr).
\end{equation}
In the calculation of the neutrino cross section, we need to evaluate,
\begin{equation}
|{\cal M}_{\nu' n',\nu n}|^{2}= \frac{1}{2} G_F^{2} \, l^{\mu\alpha} H_{\mu\alpha},
\label{matrixcuad}
\end{equation}
where $l^{\mu\alpha}$ and $H_{\mu\alpha}$ are the leptonic and hadronic traces, respectively. Now we analyze each trace separately.

\subsection{Leptonic trace}
The leptonic trace is:
\begin{eqnarray}
l^{\mu\alpha}&=&(\overline{u}_{\nu'}\gamma^\mu \frac{1}{2}\left(1-\gamma_5 \right)u_{\nu})^\dagger
(\overline{u}_{\nu'}\gamma^\alpha \frac{1}{2}\left(1-\gamma_5 \right)u_{\nu}).
\end{eqnarray}
Using standard properties of the gamma matrices, the adjoint factor of this trace can be expressed as,
\begin{eqnarray}
\frac{1}{2}\left(\overline{u}_{\nu'}\gamma^\mu \left(1-\gamma_5 \right)u_{\nu}\right)^\dagger &=&\frac{1}{2}u^\dagger_\nu \left(\gamma^\mu \left(1-\gamma_5 \right)\right)^\dagger \overline{u}^\dagger_{\nu'}=\frac{1}{2}\overline{u}_{\nu}\gamma^{\mu}\left(1-\gamma_5\right)u_{\nu'},
\end{eqnarray}
in this way, we have,
\begin{eqnarray}
l^{\mu\alpha}&=&\frac{1}{4}\overline{u}_{\nu} \gamma^{\mu} \left(1-\gamma_5\right)u_{\nu'}\overline{u}_{\nu'}\gamma^\alpha \left(1-\gamma_5 \right)u_{\nu}.
\end{eqnarray}
By using the completeness relation $u(p,s)\overline{u}(p,s)=\slashed{p}+m$, and neglecting the neutrino mass, we can write,
\begin{eqnarray}
l^{\mu\alpha}&=&\frac{1}{4}tr(\gamma^\mu \left(1-\gamma_5\right)\slashed{p}_{\nu'} \gamma^\alpha \left(1-\gamma_5 \right) \slashed{p}_\nu )=\frac{1}{2} tr( \gamma^{\mu}\slashed{p}_{\nu'}\gamma^\alpha \slashed{p}_\nu+\gamma_5 \gamma^{\mu}\slashed{p}_{\nu'}\gamma^\alpha \slashed{p}_\nu).
\end{eqnarray}
After some algebra, we found,
\begin{eqnarray}\label{}
l^{\mu\alpha}=2( p_{\nu'}^\mu p_{\nu}^\alpha + p_{\nu}^\mu p_{\nu'}^\alpha-g^{\mu\alpha} \left(p_{\nu}\cdot p_{\nu'}\right)-i\epsilon^{\mu\alpha\gamma\lambda}p_{\nu'  \gamma}p_{\nu \lambda}).
\end{eqnarray}

\subsection{Hadronic trace}
We follow similar steps as in the case of the leptonic trace, but the evaluation is more complex as we have to distinguish two terms, according to the
spin projection of the neutron. The required hadronic trace, is then,
\begin{eqnarray}
H^{s}_{\mu\alpha}&=&\left(\overline{u}_{n'}\gamma_\mu \left(C_V-C_A\gamma_5 \right) \Lambda_{s} u_{n}\right)^\dagger \left(\overline{u}_{n'}\gamma_\alpha\left(C_V-C_A\gamma_5 \right)\Lambda_{s} u_{n} \right),
\end{eqnarray}
where, as stated in the main text, we have introduced the spin projection operator over the initial neutron as,
$\Lambda_{s}=\frac{1}{2}\left(1+ \gamma_5\slashed{w}_s\right)$, where $w^s=\left(0,0,0,s \right)$ with $s=+1$ ($-1$) for spin
up (down). We re-write the adjoint factor of the hadronic trace as,
\begin{eqnarray}
\left(\overline{u}_{n'}\gamma_\mu \left(C_V-C_A\gamma_5 \right) \Lambda_{s} u_{n}\right)^\dagger&=&\frac{1}{2}\overline{u}_n \gamma^0  \left(1+ \gamma_5\slashed{w}_s\right)^\dagger \left(\gamma_\mu \left(C_V-C_A\gamma_5 \right)\right)^\dagger \gamma^0 u_{n'},
\end{eqnarray}
by making the substitution,
\begin{equation}
\gamma^0 \left(1+ \gamma_5\slashed{w}_s\right)^\dagger \left(\gamma_\mu \left(C_V-C_A\gamma_5 \right)\right)^\dagger \gamma^0=\left( 1+\gamma_5 \slashed{w}_s\right)\left(C_V+C_A \gamma_5 \right)\gamma_\mu \, ,
\end{equation}
we have,
\begin{eqnarray}
 H^{s}_{\mu\alpha}&=&\frac{1}{4}\overline{u}_n \left(1+\gamma_5 \slashed{w}_s\right)\left(C_V+C_A \gamma_5 \right)\gamma_\mu u_{n'}\overline{u}_{n'}\gamma_\alpha \left(C_V-C_A\gamma_5 \right)\left(1+ \gamma_5\slashed{w}_s\right)u_n\nonumber \\
&=&\frac{1}{4} tr(\left( 1+\gamma_5 \slashed{w}_s\right)\left(C_V+C_A \gamma_5 \right)\gamma_\mu(\slashed{p}_{n'}+m_N)\gamma_\alpha \left(C_V-C_A\gamma_5 \right)
\nonumber \\
& \times & \left(1+ \gamma_5\slashed{w}_s\right)(\slashed{p}_n +m_N)).
\end{eqnarray}
For convenience, we split this trace into three contributions: one proportional to $C^2_V$, the other one to $C^2_A$ and the last one to $C_V C_A$. After
some algebra we get,
\begin{eqnarray}
H_{\mu\alpha}^{s, \, V}&=&C^2_V\biggl(
\left(p_{n'\mu}p_{n\alpha}+p_{n'\alpha}p_{n\mu}-
g_{\mu\alpha}(p_{n'}\cdot p_n) +m_N^2
g_{\mu\alpha}\right)\left(1-w^{\beta} w_{\beta}\right)+2im_N
\epsilon_{\mu\alpha\gamma\lambda}p_{n'}^\gamma w^\lambda
\nonumber \\ &+&2i m_N \epsilon_{\mu\alpha\gamma\lambda}
w^\gamma p^\lambda_n+2w_\alpha p_{n'\mu}(w \cdot p_n)+2w_\mu
p_{n' \alpha}(w \cdot p_n)- 2 g_{\mu\alpha}(p_{n'} \cdot w)
(w\cdot p_n)\biggr),\nonumber \\ %
H_{\mu\alpha}^{s, \, A}&=&-C_A^2\biggl( -\left(p_{n' \mu}p_{n
\alpha}+p_{n' \alpha}p_{n \mu}-g_{\mu\alpha}\left(p_{n'}\cdot
p_n \right)-g_{\mu\alpha}m_N^2\right)\left(1-w^{\beta} w_{\beta}\right)+2im_N
\epsilon_{\mu\alpha\gamma\lambda}w^\gamma p_{n}^\lambda
\nonumber\\ &+& 2im_N \epsilon_{\mu\alpha\gamma\lambda}
w^\gamma p_{n'}^\lambda-2w_{\alpha}p_{n'\mu}\left(w \cdot p_{n}
\right)+ 2g_{\mu\alpha}\left(w \cdot p_{n'}\right)\left(w \cdot
p_{n} \right)-2w_{\mu}p_{n' \alpha}\left(w\cdot
p_{n}\right)\biggr),\nonumber \\ H_{\mu\alpha}^{s, \, VA} &=&C_V
C_A\biggl(-4m_N \left(p_{n'\mu} w_{\alpha}+ p_{n'\alpha}
w_{\mu}-g_{\mu\alpha}\left( p_{n'} \cdot w \right)
\right)-4i\epsilon_{\mu\alpha\gamma\lambda} p_{n'}^\gamma
p_n^\lambda-3i\epsilon_{\mu\alpha\gamma\lambda} p_{n'}^\gamma
p_n^\lambda w^{\beta} w_{\beta} \nonumber \\
&+&4i\epsilon_{\mu\gamma\lambda\rho}w_{\alpha}w^\gamma
p_{n'}^\lambda
p_n^\rho-4i\epsilon_{\alpha\gamma\lambda\rho}w_\mu w^\gamma
p_{n'}^\lambda
p_n^\rho+4i\epsilon_{\mu\alpha\gamma\lambda}\left(p_{n'} \cdot
w \right) w^\gamma p_n^\lambda\biggr),
\label{hadront}
\end{eqnarray}
where for simplicity we have omitted the spin index in all $w$.
Obviously we have,
\begin{equation}
H^{s}_{\mu\alpha}=H_{\mu\alpha}^{s, \, V}+H_{\mu\alpha}^{s, \, A}+H_{\mu\alpha}^{s, \, VA}.
\end{equation}
It is convenient to simplify these expressions by neglecting the neutron momenta, for both the incoming and the outgoing neutron. In this case,
we have, $ \left(p_{n'} \cdot w \right) \cong 0$ and $\left(w \cdot p_{n} \right)\cong 0$, and Eqs.~(\ref{hadront}), reduced to,
\begin{eqnarray}
H_{\mu\alpha}^{s, \, V}&=& C^2_V \left((p_{n'\mu}p_{n\alpha}+p_{n' \alpha}p_{n\mu}-g_{\mu\alpha}(p_{n'} \cdot p_{n})+m_N^2g_{\mu\alpha})(1-w^{\beta} w_{\beta})+2im_N \epsilon_{\mu\alpha\lambda\gamma} w^\lambda(p_n^\gamma-p_{n'}^\gamma)\right),\nonumber \\
H_{\mu\alpha}^{s, \, A}&=&C_A^2\left((p_{n' \mu}p_{n\alpha}+p_{n' \alpha}p_{n\mu}-g_{\mu\alpha}\left(p_{n'}\cdot p_{n}\right)-g_{\mu\alpha}m_N^2)(1-w^{\beta} w_{\beta})-2im_N \epsilon_{\mu\alpha\lambda\gamma}w^\lambda(p_n^\gamma+p_{n'}^\gamma)\right),\nonumber \\
H_{\mu\alpha}^{s, \, VA} &=&-4m_N C_V C_A (p_{n' \mu} w_{\alpha}+ p_{n' \alpha} w_{\mu}).
\end{eqnarray}
Finally, we now build up the spin up and down components, that is $H_{\mu\alpha}^+$ and $H_{\mu\alpha}^-$, as,
\begin{eqnarray}
H_{\mu\alpha}^+&=&\biggl(C_V^2\left( 2\left( p_{n' \mu}p_{n\alpha}+p_{n'\alpha}p_{n\mu}- g_{\mu\alpha} (p_{n'} \cdot p_n) + m_N^2 g_{\mu\alpha}\right)+2im_N\epsilon_{\mu\alpha\lambda z}\left(p_{n'}^\lambda-p_n^\lambda\right) \right)\nonumber\\
&+&C^2_A \left( 2\left( p_{n' \mu} p_{n \alpha}+p_{n' \alpha} p_{n\mu}- g_{\mu\alpha}(p_{n'} \cdot p_n)-m^2_N g_{\mu\alpha}\right)+2im_N \epsilon_{\mu\alpha\gamma z}\left(p_{n'}^\gamma +p_{n}^\gamma\right)\right)\nonumber\\
&-&4C_V C_A m_N(p_{n' \mu} g_{\alpha z}+g_{\mu z} p_{n' \alpha})\biggr)
\end{eqnarray}
and
\begin{eqnarray}
H_{\mu\alpha}^-&=&\biggl( C_V^2 \left(2 \left( p_{n' \mu}p_{n \alpha}+p_{n' \alpha}p_{n\mu}- g_{\mu\alpha} (p_{n'} \cdot p_n) + m_N^2 g_{\mu\alpha}\right)-2im_N\epsilon_{\mu\alpha\lambda z}\left(p_{n'}^\lambda-p_n^\lambda\right) \right)\nonumber\\
&+&C^2_A \left( 2\left(p_{n' \mu} p_{n \alpha}+p_{n' \alpha} p_{n \mu}- g_{\mu\alpha}(p_{n'} \cdot p_n)-m^2_N g_{\mu\alpha}\right)-2im_N \epsilon_{\mu\alpha\gamma z}\left(p_{n'}^\gamma +p_{n}^\gamma \right)\right)\nonumber\\
&+&4C_V C_A m_N\left(p_{n'\mu} g_{\alpha z}+g_{\mu z} p_{n'\alpha}\right)\biggr).
\end{eqnarray}
Note that we have used the index $z$ for the third spatial component of the four vectors.

\subsection{Evaluation of $|{\cal M}_{\nu' n',\nu n}|^2$}
From Eq.~(\ref{matrixcuad}), we contract the leptonic and hadronic traces to build up $|{\cal M}_{\nu' n',\nu n}|^2$, which is
also divided into a spin up and a spin down contribution. In the following expressions, we have added the spin asymmetry factor ${\cal A}$
(see Eq.~(\ref{assim})).
\begin{eqnarray}
\biggl|\mathcal{M}_{\nu'n',\nu n}^+\biggr|^2&=&8G_F^2 \, \frac{1+{\cal A}}{2} \, \left( p_{\nu'}^\mu p_\nu^\alpha+ p_\nu^\mu p_{\nu '}^\alpha-g^{\mu\alpha} \left( p_{\nu '} \cdot p_\nu \right)-i\epsilon^{\mu\alpha\gamma\lambda}p_{\nu'\gamma}p_{\nu\lambda}\right)\nonumber\\
& \times &\biggl(C_V^2\left(\left( p_{n'\mu}p_{n\alpha}+p_{n'\alpha}p_{n\mu}- g_{\mu\alpha} (p_{n'} \cdot p_n) + (m^{*}_{+})^2 g_{\mu\alpha}\right)+i m^{*}_{+} \epsilon_{\mu\alpha\rho z}\left(p_{n'}^\rho-p_n^\rho\right) \right)\nonumber\\
&+&C^2_A \left(\left(p_{n' \mu} p_{n\alpha}+p_{n' \alpha} p_{n \mu}- g_{\mu\alpha}(p_{n'} \cdot p_n)- (m^{*}_{+})^2  g_{\mu\alpha}\right)+i m^{*}_{+} \epsilon_{\mu\alpha\rho z}\left(p_{n'}^\rho +p_{n}^\rho\right)\right)\nonumber\\
&-&2C_V C_A m^{*}_{+} \left(p_{n' \mu} g_{\alpha z}+g_{\mu z} p_{n' \alpha}\right) \biggr),
\end{eqnarray}
\begin{eqnarray}
\biggl|\mathcal{M}_{\nu'n',\nu n}^-\biggr|^2&=&8G_F^2 \, \frac{1-{\cal A}}{2} \, \left( p_{\nu'}^\mu p_{\nu}^\alpha+ p_{\nu}^\mu p_{\nu'}^\alpha-g^{\mu\alpha}\left(p_{\nu'}\cdot p_{\nu} \right)-i\epsilon^{\mu\alpha\gamma\lambda}p_{\nu'\gamma}p_{\nu \lambda}\right)\nonumber\\
&\times&\biggl(C_V^2 \left(\left( p_{n' \mu}p_{n\alpha}+p_{n' \alpha}p_{n\mu}- g_{\mu\alpha}(p_{n'}\cdot p_n) + (m^{*}_{-})^2 g_{\mu\alpha}\right)-i m^{*}_{-} \epsilon_{\mu\alpha\rho z}\left(p_{n'}^\rho-p_n^\rho\right)\right)\nonumber\\
&+&C^2_A \left( \left( p_{n'\mu}p_{n \alpha}+p_{n' \alpha} p_{n \mu}- g_{\mu\alpha}(p_{n'} \cdot p_n)- (m^{*}_{-})^2 g_{\mu\alpha}\right)-i m^{*}_{-}  \epsilon_{\mu\alpha\rho z}\left(p_{n'}^\rho + p_{n}^\rho\right)\right)\nonumber\\
&+&2C_V C_A m^{*}_{-} \left(p_{n' \mu} g_{\alpha z}+g_{\mu z} p_{n' \alpha}\right) \biggr),
\end{eqnarray}
where we have replaced the bare mass $m_N$, by the effective mass $m^{*}_{+}$
($m^{*}_{-}$) for neutrons with spin up (down).
Finally, by using
$\epsilon^{\xi\phi \gamma\nu} \epsilon_{\lambda\rho \gamma\nu}=- 2\left(\delta^\xi_\lambda \delta^\phi_\rho - \delta^\xi_\rho \delta^\phi_\lambda\right)$,
we obtain the Eqs.~(\ref{mreld}) and (\ref{mrelu}).
\newpage
\section*{Acknowledgements}
This work was partially supported by the
CONICET, Argentina, under contract PIP00273, and by ``PHAROS: The multi-messenger physics and astrophysics of neutron stars", COST Action CA16214.


\newpage

\end{document}